\documentclass[compsoc, conference, a4paper, 10pt, times]{IEEEtran}
\IEEEoverridecommandlockouts
\usepackage{wrapfig}
\usepackage{amsmath}
\usepackage{textcomp}
\usepackage{xcolor}
\usepackage{soul}
\usepackage{braket}
\usepackage{algorithmic}
\usepackage[ruled,vlined, linesnumbered]{algorithm2e}
\usepackage{graphicx}
\usepackage{cite}
\usepackage{amsmath,amssymb,amsfonts}
\usepackage{algorithmic}
\usepackage{graphicx}
\usepackage{textcomp}
\usepackage{xcolor}
\def\BibTeX{{\rm B\kern-.05em{\sc i\kern-.025em b}\kern-.08em
    T\kern-.1667em\lower.7ex\hbox{E}\kern-.125emX}}
\begin{document}

\title{
Trustworthy Computing using Untrusted Cloud-Based Quantum Hardware\\}

\author{\IEEEauthorblockN{Suryansh Upadhyay}
\textit{The Pennsylvania State University}\\
PA,USA \\
sju5079@psu.edu
\and


\and

\IEEEauthorblockN{Swaroop Ghosh}
\textit{The Pennsylvania State University}\\
PA,USA \\
szg212@psu.edu

}

\maketitle

\begin{abstract}

Security and reliability are primary concerns in any computing paradigm, including quantum computing. Currently, users can access quantum computers through a cloud-based platform where they can run their programs on a suite of quantum computers. 
As the quantum computing ecosystem grows in popularity and utility, it is reasonable to expect that more companies including untrusted/less-trusted/unreliable vendors \footnote{untrusted, less-trusted and unreliable vendors are used interchageabily in this paper} will begin offering quantum computers as hardware-as-a-service at varied price/performance points. Since computing time on quantum hardware is expensive and the access queue could be long, the users will be motivated to use the cheaper and readily available but unreliable/less-trusted hardware. 
The less-trusted vendors can 
tamper with the results, providing a sub-optimal solution to the user. For applications such as, critical infrastructure optimization, the inferior solution may have significant socio-political implications. Since quantum computers cannot be simulated in classical computers, users have no way of verifying the computation outcome. In this paper, we model this adversarial tampering and simulate it's impact on a number of pure quantum and hybrid quantum classical workloads. To guarantee trustworthy computing for a mixture of trusted and untrusted hardware, we propose distributing the total number of shots (i.e., number of repeated execution of a quantum program for computation) equally among the various hardware options. On average, we note $\approx$ 30X and $\approx$ 1.5X improvement across the pure quantum workloads and a maximum improvement of $\approx$ 5X for hybrid-classical algorithm in the chosen quality metrics. We also propose an intelligent run adaptive split heuristic leveraging temporal variation in hardware quality to user's advantage, allowing them to identify tampered/untrustworthy hardware at runtime and allocate more number of shots to the reliable hardware, which results in a maximum improvement of $\approx$ 190X and $\approx$ 9X across the pure quantum workloads and an improvement of up to $\approx$ 2.5X for hybrid-classical algorithm.

\end{abstract}


\section{Introduction}

Quantum computing (QC) can solve many combinatorial problems exponentially faster than classical counterparts by leveraging superposition and entanglement properties. Examples include machine learning \cite{b1}, security \cite{b2}, drug discovery \cite{b3}, computational quantum chemistry \cite{b4} and optimization \cite{b20}. 
However, quantum computing faces technical challenges like quantum bit (qubit) decoherence, measurement error, gate errors and temporal variation. As a result, a quantum computer may sample the wrong output for a specific quantum circuit. While quantum error correction codes (QEC) can provide reliable operations \cite{b5}, they require thousands of physical qubits per logical qubit, making them impractical in the foreseeable future. 
Existing Noisy Intermediate-Scale Quantum (NISQ) computers have a few hundred qubits and operate in the presence of noise. The NISQ computing paradigm offers hope to solve important problems such as, discrete optimization and quantum chemical simulations.
Since noisy computers are less powerful and qubit limited, various hybrid algorithms are being pursued, such as, the Quantum Approximate Optimization Algorithm (QAOA) and Variational Quantum Eigensolver (VQE), in which a classical computer iteratively drives the parameters of a quantum circuit. The purpose of the classical computer is to tune the parameters that will guide the quantum program to the best solution for a given problem. On high-quality hardware with stable qubits, the algorithm is likely to converge to the optimal solution faster, i.e., with fewer iterations.
Existing quantum computers also have limited connectivity and can only execute a small set of native instructions. As a result, a high-level quantum program is compiled to meet the connectivity and native instruction limitations.

\begin{figure}
    \centering
    \includegraphics[width= 3.25in]{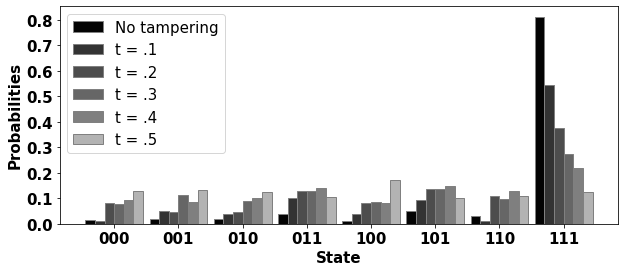}
    \caption{Sample benchmark (toffoli$_-$n3, correct output = `111') simulated on the fake back-end (Fake$_-$montreal) for 10,000 shots. Changing the tampering coefficient models the extent of adversarial tampering (t). For t =0.5 erroneous state `100' becomes the most occurring output.
    }
    \label{1}
\vspace{-4mm}
\end{figure}



Security and reliability are primary concerns in quantum computing. Researchers are currently exploring a suite of quantum computers offered by IBM, Rigetti, IonQ, and D-Wave (via a cloud-based platform) to solve optimization problems. The hardware vendors of quantum computers provide a compiler for their hardware, such as, IBM's Qiskit compiler \cite{b6}, Rigetti's QuilC compiler \cite{b7}, and so on. Users can create circuits for specific hardware and upload them to the cloud, where they are queued. The results of the experiment are returned to the user once the experiment is completed.

As the quantum computing ecosystem evolves, third-party service providers are expected to emerge offering potentially higher performance. This will entice users to utilize these services. For example, some third-party compilers like Orquestra \cite{b8} and tKet \cite{b9}, are appearing that support hardware from multiple vendors. Baidu, the Chinese internet giant, recently announced an ``all-platform quantum hardware-software integration solution that provides access to various quantum chips via mobile app, PC, and cloud.'' referred to as ``Liang Xi'' \cite{b38}. It provides flexible quantum services via private deployment, cloud services, and hardware access, and can connect to other third-party quantum computers. These trends not only lead to reliance on untrustworthy third-party compilers and hardware suites instead of trusted counterparts, but also to reliance on third-party service providers, which can pose a security risk.


\textbf{Proposed attack model:} In this paper, we discuss a security risk associated with the use of third-party service providers and/or any untrusted vendor. In the proposed attack model, less-trusted quantum service providers can pose as trustworthy and tamper with the results, resulting in the worst-case scenario of users receiving a sub-optimal solution. To show the extent of damage done by the proposed tampering model, we run a simple program (with a known solution) on tampered and non-tampered hardware and compare the probability distributions of basis states for both cases (Fig. \ref{1}). The correct output is `111'. The tampering coefficient (t) models the various degrees of hardware tampering. As t increases, the probability of basis state `111' decreases while the probabilities of the other erroneous states increase. For the case of t=0.5, state `111' is no longer the dominant output; instead, the incorrect state `100' becomes the dominant which can be reported back to the user. In practical scenario, as the correct solution to the optimization problem is unknown, the user must rely on the sub-optimal output of the tampered quantum computer.

\textbf{Novelty:} In the classical domain, existence of malicious third-party cloud computers and associated vulnerabilities are well known such as, hacking, data breaches and insecure APIs \cite{b40}. Therefore, as the quantum computing ecosystem grows in popularity and utility, it is reasonable to expect that more companies including untrustworthy third-party vendors will begin offering quantum computers as hardware-as-a-service posing a variety of security challenges. Although the proposed attack model sound similar to classical domain, quantum computing bring new twists e.g., (a) users can not verify the results (which is possible in classical domain) after adversarial tampering since the correct output of a quantum program cannot be computed in classical computer, (b) the results of computation are probability distribution of basis states (instead of deterministic results in classical domain) which opens up new ways of tampering via manipulation of basis state probabilities, (c) the attack models could be low-overhead (e.g., by manipulation of gate error rates) which can be challenging to detect, while significantly affecting the probability of program's correct output.

\textbf{Viability of the proposed attack model:} The proposed attack model is feasible since, (a) Quantum computers are expensive. Customers can purchase quantum computing services over the cloud. We investigated the prices charged by AWS Braket for access to IonQ, OQC, and Rigetti quantum processors, IBM for access to their own processors, and Google Cloud for access to IonQ's quantum computer. Assuming 1ms runtime per shot, current prices range from $\$$0.35 to $\$$1.60 per second for qubit counts in the range of 8 to 40. To factor a 2048 bit product of two primes, for example, a quantum computer will need approximately 25 billion operations in 14238 logical qubits \cite{b37}, which equals 432 billion qubit-seconds. With many new vendors entering the quantum service market, it is likely that some of these vendors will be untrusted who will offer access to quantum hardware via the cloud at a lower cost, enticing users to use their services. This is more likely if the third party is based offshore, where labor, fabrication, and packaging costs are cheaper. (b) Access to quantum computers incur long wait time. When a user submits a job to a quantum system, it enters the scheduler where it is queued. The rapid expansion of quantum computing requirements has worsened the competition for these already scarce resources. For IBM Quantum systems, \cite{b39} reports that only about 20$\%$ of total circuits have ideal queuing times of less than a minute. The average wait time is about 60 minutes. Furthermore, more than 30$\%$ of the jobs have queuing times of more than 2 hours, and 10$\%$ of the jobs are queued for as long as a day or longer! Third party vendors may provide access to quantum hardware with little or no wait time. Quick access may be vital for quantum machine learning applications to lower the training and inference time. 

 \begin{figure}
    \centering
    \includegraphics[width= 3.25in]{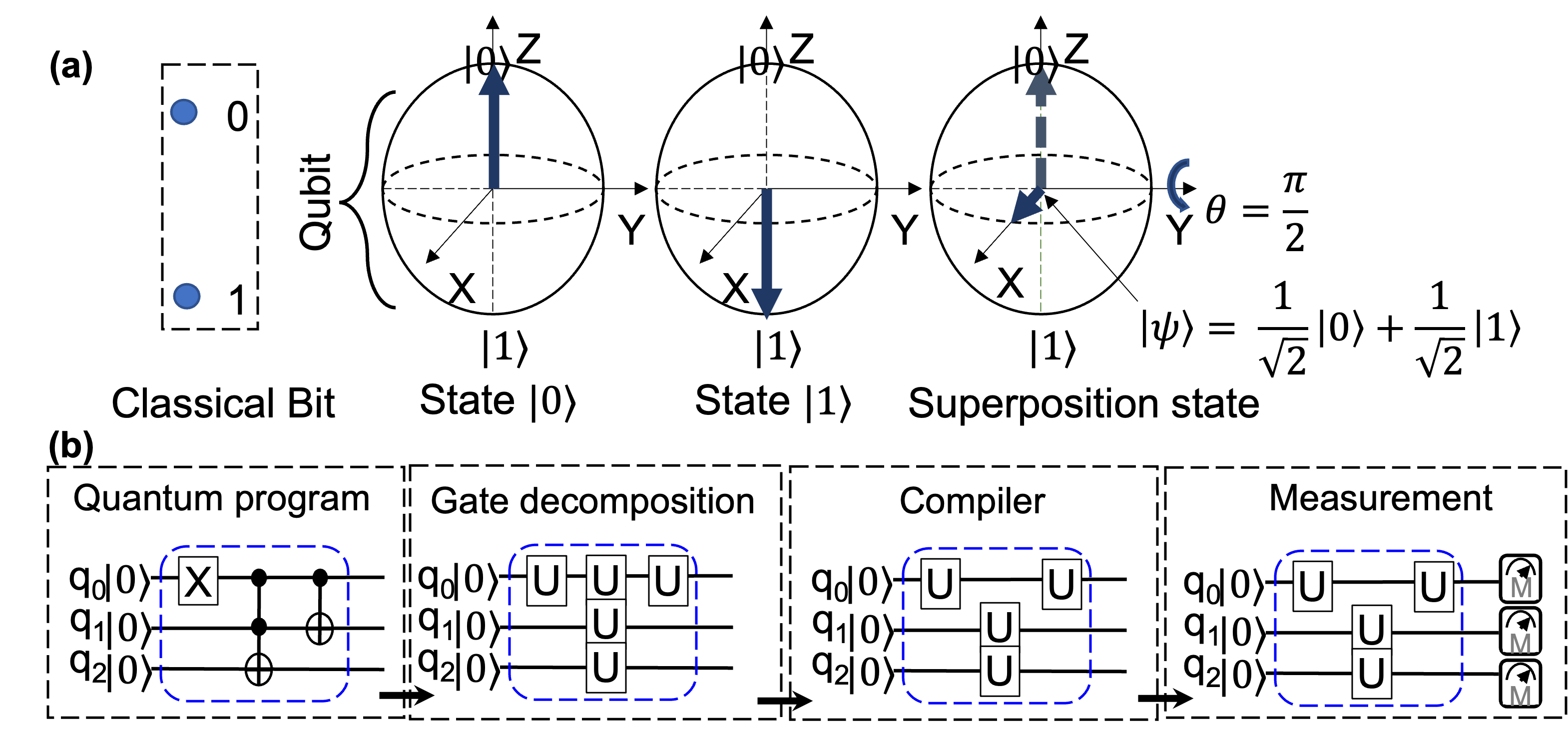}
    \caption{a) Classical bit compared to a qubit. Bloch sphere representation of state $\ket{0}$, state $\ket{1}$ and a superposition state. b) Overview of quantum program flow. A quantum program is a set of quantum gates. Before the compiler maps the qubits to target hardware, the gates are decomposed to hardware supported native gates (U $\in$ supported gates), and circuit optimization (e.g., redundant gates are removed) is also performed before the program is run, followed by measurement operation.}
    \label{2}
\vspace{-4mm}
\end{figure}

\textbf{Proposed solution:} We propose two solutions, (a) Split and distribution of shots/trials: Even though qubits process $\ket{0}$ and $\ket{1}$ together, measurement snaps the qubit to classical 0 or 1. To measure the qubit state which could be a mixture of 0 and 1, traditionally the program is executed multiple times (called \emph{shots}) on the hardware. To mitigate the adversarial tampering we propose splitting shots on available hardware. The idea is to distribute the computation among the various hardware (a mix of trusted and untrusted ones or mixture of untrusted hardware for multiple vendors) available. The results from individual hardware and iterations are stitched or combined to obtain the probability distribution of the solution space. (b) Intelligent shot/trial split and distribution: Although splitting of shots is effective, users may end up using trusted and untrusted hardware equally which may not be optimal in terms of performance. We propose an intelligent run-adaptive shot distribution which leverages temporal variation in hardware quality to identify untrusted hardware and bias the number of shots to favor trusted/reliable hardware to maximize the overall computation quality. 
\textbf{Novelty:} (a) Redundant computation for resilience is well-known in classical domain. However, the proposed approach in quantum domain avoids performing any redundant computation by keeping the total number of trails/shots (that is needed to estimate the basis state probability of the circuit) fixed at original value while improving the resilience to tampering. Distribution of shots to multiple vendors/hardware may increase the overall expense only if a hardware with higher price than the baseline hardware is employed for shot distribution. (b) The proposed approach of identification of tampered hardware at run-time by monitoring the dynamic behavior of computation results is novel and specific to quantum domain. (c) The proposed attack model and defenses are evaluated for two prominent computation models namely, pure quantum and hybrid quantum-classical approach.   

\textbf{Research challenges:} Although the proposed tampering model and shot distribution based defense may appear trivial, there are many associated technical challenges. For example, (a) what should be the tampering approach? Should all qubits be tampered equally or randomly or selectively and by how much to evade detection? (Section 4.4) (b) how to decide the split boundary i.e., equal or asymmetric split? (Section 6.2) (c) since the trustworthiness of the hardware is not known in advance, how can the user distribute the shots to maximize the quality of solution? (Section 6.3) (d) what kind of metric should be used to evaluate the impact of tampering and effectiveness of the defense? (Section 5.4) (e) does the tampering affect all quantum algorithms equally? We address such research questions in this paper through extensive analysis (Section 6.4).   

\textbf{Contributions:} (1) We propose and compare random vs. selective tampering model for untrustworthy third-party hardware vendors. (2) To counteract adversarial tampering, we propose equally distributing shots among available hardware and an intelligent run-adaptive shot splitting heuristic leveraging temporal variation. (3) We demonstrate the effectiveness of our proposed approach for pure quantum and hybrid quantum-classical workloads on a variety of fake back-ends. (4) We validate the attack model and the proposed defense on real hardware. 
 
In the remaining paper, Sections II and III provide quantum computing background and related work, respectively. The proposed attack model is described in Section IV. Section V proposes the tampering model, simulations and evaluation on real hardware. Section VI presents and evaluates the defense using simulations and experiments on real hardware. Section VII concludes the paper.

\section{Background}

In this section, we discuss the basics of a quantum computer, quantum service providers, quantum security and the terminologies used in this paper.

\subsection{Qubits}

Qubits are the building blocks of a quantum computer that store data as various internal states (i.e., $\ket{0}$ and $\ket{1}$). In contrast to a classical bit, which can only be either 0 or 1, a qubit can concurrently be in both $\ket{0}$ and $\ket{1}$ due to quantum superposition. Hence, while a standard n-bit register can only represent one of the $2^n$ basis states, an n-qubit system can represent all $2^n$ basis states concurrently. A qubit state is represented as  $\varphi$ = a $\ket{0}$ + b $\ket{1}$ where a and b are complex probability amplitudes of states $\ket{0}$ and $\ket{1}$ respectively. The qubit is reduced to a single state by measurement, i.e., a pure state of $\ket{0}$ or $\ket{1}$  with probability of $|a|^2$ and $|b|^2$ respectively. Therefore, a quantum program is executed multiple times (also called \emph{shots}) to obtain the probability of basis states. 
Qubits are frequently visualized as a point on the so-called Bloch Sphere Fig. \ref{2}a. This representation shows both the phase and the probabilities of measuring a qubit as either of the basis states, which are represented as the north and south poles of the sphere.


\subsection{Quantum Gates}

The gate operations change the amplitudes of the qubits to produce the desired output. A quantum program executes a series of gate operations (using laser pulses in ion trap qubits and RF pulses in superconducting qubits) on a group of correctly initialized qubits. 
Fig. \ref{2}b depicts a high-level overview of the quantum ecosystem. Mathematically, quantum gates are represented using unitary matrices (a matrix U is unitary if U$U^\dagger$ = I, where U$^\dagger$ is the adjoint of matrix U and I is the identity matrix). For an n-qubit gate, the dimension of the unitary matrix is 2n×2n. Any unitary matrix can be a quantum gate. Only a few gates, known as the quantum hardware's native gates 
are currently practical in current systems. ID, RZ, SX, X (single qubit gates), and CNOT (2-qubit gate) are the basic gates for IBM systems. 
All complex gates in a quantum circuit are first decomposed to native gates.

\begin{figure}
    \centering
    \includegraphics[width= 3.25in]{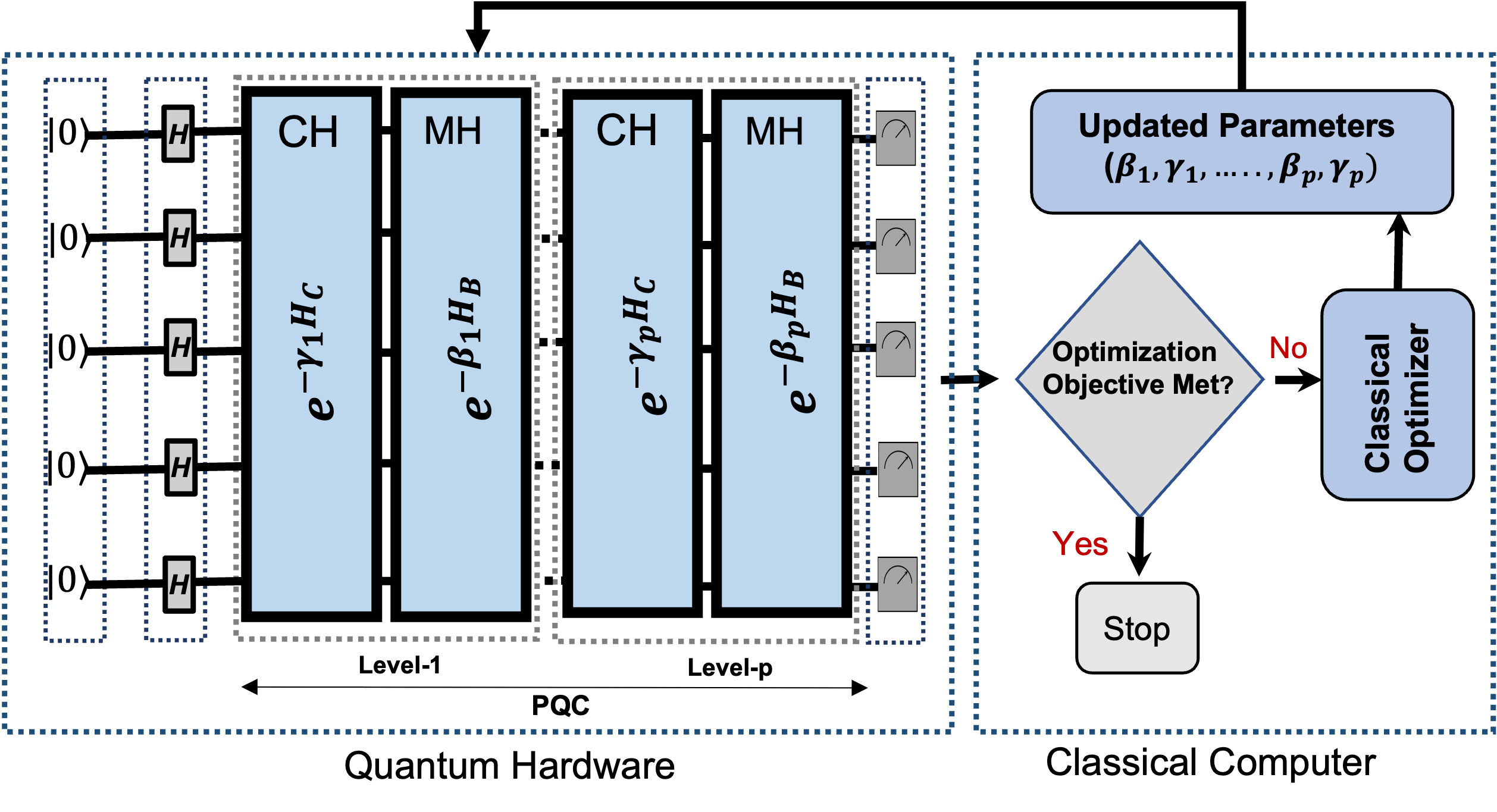}
    \caption{Schematic of a p-level quantum-classical hybrid algorithm QAOA. A quantum circuit takes input qubit states and alternately applies Cost Hamiltonian (CH) and Mixing Hamiltonian (MH) `p' times and the final state is measured to obtain expectation value with respect to the objective function. This is fed to a (classical) optimizer to find the best parameters$(\gamma, \beta)$ that maximizes the cost.
    }
    \label{16}
\vspace{-4mm}
\end{figure}

\subsection{Quantum Program and Instruction Sets}

A quantum program can be represented in the well-adopted quantum circuit model \cite{b29}. 
It is similar to classical computing in that it may be characterized as a set of quantum gates operating on qubits to converge the output to a particular solution. A quantum program is composed of logical qubit variables and quantum operations that can modify the state of the qubits. Higher-level algorithms are converted into physical instructions that can be executed on quantum processors using quantum instruction sets. Various instruction set architectures are available e.g., cQASM, OpenQASM, Quil, Blackbird etc.

\subsection{Quantum Error}

 Quantum gates are realized with pulses that can be erroneous. For example, consider the R$_{Y}(\pi/2$) gate \emph{(Ry gate is a single-qubit gate that rotates the state of a qubit around the y-axis of the Bloch sphere by a certain angle)}. Due to variation, the pulse intended for a $\pi/2$ rotation may under-rotate or over-rotate, leading to erroneous logical operation.
Quantum gates are also prone to error due to noise and decoherence \cite{b10}. Hence qubits interact with their surroundings and lose their states making the output of a quantum circuit error-prone. The deeper quantum circuit needs more time to execute and gets affected by decoherence which is usually characterized by the relaxation time (T1) and the dephasing time (T2). The buildup of gate error \cite{b11} is also accelerated by more gates in the circuit. Cross-talk is another form of quantum error where parallel gate operations on several qubits can negatively impact each others performance. Because of measurement circuitry imperfections, reading out a qubit containing a 1 may result in a 0 and vice versa. 
The qubit quality metrics e.g., gate error, measurement error, decoherence/dephasing and cross-talk vary significantly over time \cite{b31}. A program running on quantum hardware may not always exhibit the repeatable behavior due to temporal variation. This also accounts for the hardware converging to a different outcome for the same program at different points in time. For example, assuming the qubits are prepared in the basis state $\ket{00}$, and ideally, the qubits should converge to $\ket{10}$ state at the end of the execution period. A projective measurement on the target hardware is expected to generate a measurement of $\ket{10}$ most of the time. However, due to temporal variations of the qubit quality metrics, we receive different outcomes at different points in time.

\subsection{Hardware Variability}

Hardware variability manifests itself in quantum computing as variation in \emph{hardware performance}, or more precisely, different gate error rates, decoherence times and so on across quantum devices. 

\subsection{Cloud-based Quantum Backends}

Bundles of tools known as quantum software development kits enable users to create and modify quantum applications. 
Currently, IBM, Google, Microsoft, Qutech, QC Ware and AWS Braket are some of the top vendors that provide access to quantum hardware (both superconducting and Trapped Ion qubits) to the users over cloud. However, access to quantum computers is expensive motivating the users to explore less expensive quantum hardware to solve their problems. In future, quantum computers could be available from less-trusted third party (that may be located offshore). This situation is very similar to untrusted compilers that may produce very optimal circuit but may steal sensitive intellectual property \cite{b32}. 



\subsection{Quantum Algorithms}

Quantum algorithms are programs that run on a realistic model of quantum computation (the most common being the quantum circuit model \cite{b29}) and are intrinsically quantum or employ some important aspect of quantum computation such as, quantum superposition or entanglement. 
The first quantum computing algorithms were designed with fault tolerant quantum computer in mind, with the quantum gate model studied largely without noise \cite{b30}. Shor's algorithm for factoring and Grover's algorithm for searching an unstructured database or an unordered list are two of the most well-known algorithms. 

Currently, a class of delicate, noisy, and error-prone quantum computers called Noisy Intermediate-Scale Quantum (NISQ) computers \cite{b12} are commercially available. Error correction cannot be used by NISQ computers with a few hundreds of qubits \cite{b13}, not even for a task needing just a few dozens of logical qubits. 
Due to their small qubit capacity, NISQ devices cannot effectively implement quantum algorithms like Shor's algorithm, which exemplifies the quantum computing paradigm. However, a number of quantum-classical hybrid algorithms \cite{b21, b22, b23} based on variational principles have been developed to demonstrate quantum supremacy and to make the greatest use of the currently available near-term devices. These hybrid algorithms accelerate a task above it's entirely classical counterpart by combining the power of a quantum and a classical computer.
\subsubsection{PQC}
PQC is composed of a set of parameterized and controlled single qubit gates. A classical optimizer iteratively optimizes the parameters to achieve the desired input-output relationship. A PQC is used by a quantum processor to prepare a quantum state. A classical optimizer is then fed an output distribution created by repeatedly measuring the quantum state. Based on the output distribution, the classical computer generates a new set of optimized parameters for the PQC which is then fed-back to the quantum computer. The entire procedure keeps running in a closed loop until a traditional optimization target is met. In recent years, quantum routines that are inherently resilient to errors have been developed using PQC \cite{b20, b33, b34}. 

\subsubsection{QAOA}
QAOA is a hybrid quantum-classical variational algorithm designed to tackle combinatorial optimization problems. A p-level variational circuit with 2p variational parameters creates the quantum state in QAOA. Even at the smallest circuit depth (p = 1), QAOA delivers non-trivial verifiable performance guarantees, and the performance is anticipated to get better as the p-value increases \cite{b14}. Fig. \ref{16} gives the overview of QAOA to solve a combinatorial problem. 
Recent developments in finding effective parameters for QAOA have been developed \cite{b14,b15,b16,b17}. 

In QAOA, a qubit is used to represent each of the binary variables in the target C(z). In each of the p levels of the QAOA circuit, the classical objective function C(z) is transformed into a quantum problem Hamiltonian Fig.\ref{16}. With optimal values of the control parameters, the output of the QAOA instance is sampled many times and the classical cost function is evaluated with each of these samples. The sample measurement that gives the highest cost is taken as the solution \cite{b19}. In a quantum classical optimization procedure,the expectation value of $H_C$ is determined in the variational quantum state $ E_p(\gamma,\beta)= \varphi_p(\gamma, \beta)|H_C|\varphi_p(\gamma, \beta)$. A classical optimizer iteratively updates these variables $(\gamma, \beta)$ so as to maximize $E_p(\gamma, \beta)$. A figure of merit (FOM) for benchmarking the performance of QAOA is the approximation ratio (AR) and is given as \cite{b14}
$    AR = E_p(\gamma, \beta)/Cmax$
where $Cmax = MaxSat(C(z))$.


\begin{figure*}
    \centering
    \includegraphics[width= 6.75in]{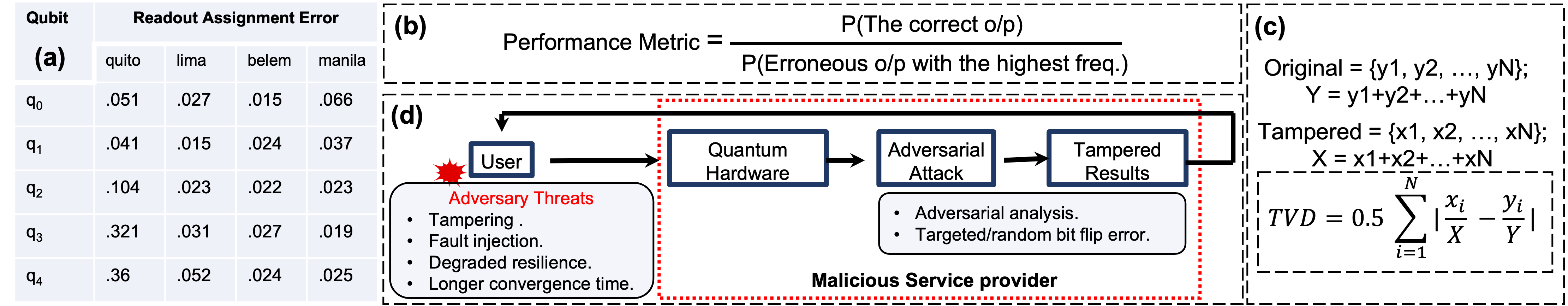}
    \caption{a) Readout assignment errors for various IBM quantum hardware. b) Performance Metric (PM). c) Total Variational Distance (TVD) between two sets of probability densities. d) Proposed attack model where the attacker introduces targeted or random tampering, resulting in the users receiving a less-than-optimal solution.
    }
    \label{3}
\vspace{-4mm}
\end{figure*}

\section{Related work}
Several recent works on the security of quantum computing \cite{b2, b28, b32, b35, b36} exist in literature. 
The authors of \cite{b35} consider an attack model where a rogue element in the quantum cloud reports incorrect device calibration data, causing a user to run his/her program on an inferior set of qubits. The changes in the qubit error rates alter the quantum circuit mapping for recently proposed policies that favor good quality qubits (to map logical to physical qubits) over poor quality ones. The resulting physical quantum circuit can incur high error rates, and thus, significantly degrade the output quality. The authors propose that test points be added to the circuit to detect any dynamic malicious changes to the calibration data. The objective of our attack model is to tamper with the result so that incorrect or sub-optimal outcome is reported to the user. As such, the impact of the proposed attack is much higher. The proposed attack is also low-overhead since it only involves manipulation of qubit outcomes post-measurement whereas \cite{b35} will require complex gate pulse manipulation to increase the error rate. 
The proposed equal shot distribution and adaptive shot splitting approach to improve resilience is also significantly different than test point insertion. 

In \cite{b28}, Ensemble of Diverse Mappings (EDM) is proposed to tolerate correlated errors and improve the NISQ machine's ability to infer the correct answer. Rather than using a single mapping for all the shots, EDM uses multiple mappings and divide their shots among these different mappings on a single piece of hardware, then merge the output to get the final solution space. However, this is yet another case of mapping agnostic optimization. If the hardware used for EDM is tampered, suboptimal solutions will be returned even with different mapping splits. In this paper, we propose heuristics to counter adversarial tampering as well as methods for detecting tampered hardware.
The attack model in \cite{b2} assumes a malicious entity in the cloud that can schedule a user circuit to inferior hardware rather than the requested one. To authenticate the requested device, they propose quantum PUFs (QuPUFs). In our attack model, users have a choice of hardware but are unaware of their trustworthiness. 

\begin{figure}
    \centering
    \includegraphics[width= 3.25in]{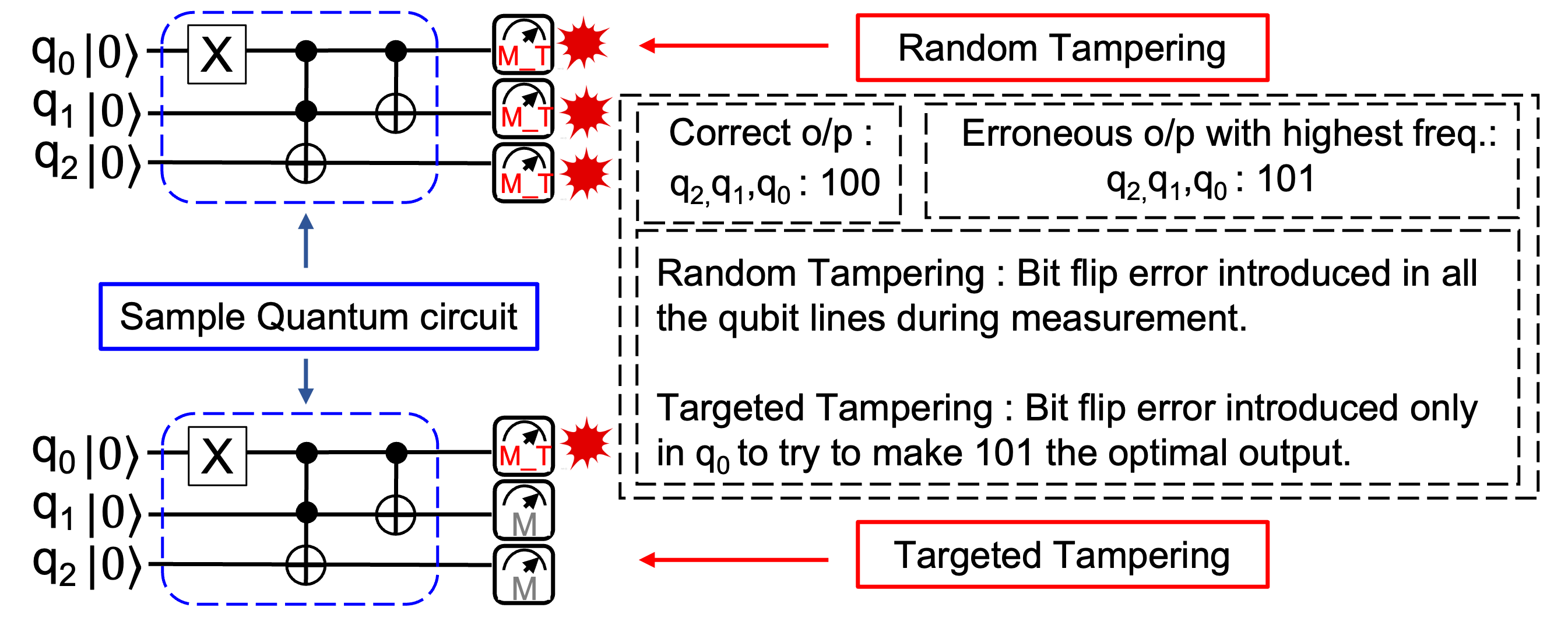}
    \caption{Adversarial random tampering and targeted tampering by introducing a bit flip error during measurement.
    }
    \label{4}
\vspace{-4mm}
\end{figure}

\section{Proposed attack model}

In this section, we describe the attack model and a few methods for the adversary to tamper with the results.

\subsection{Basic Idea}

We consider that the quantum hardware available via cloud service may tamper with the computation outcome. The objective is to manipulate the results that could have financial and/or socio-political implications. This is feasible under following scenarios, (a) untrusted third party may offer access to reliable and trusted quantum computers e.g., from IBM in the future but may tamper the computation results, 
(b) untrusted vendors may offer access to untrusted quantum hardware via cloud at cheaper price and/or quick access (without wait queue) motivating the users to avail their services. The less-trusted quantum service providers can pose as trustworthy hardware providers and inject targeted/random tampering, causing the sub-optimal solution to be sent back 
to the users. 
For both scenarios, the user will be forced to trust the less-than-ideal output from the quantum computer since the correct solution to the optimization problem is unknown. 

\subsection{Adversary Capabilities}

We assume that adversary, (a) has access to the measured results of the program run by the user. This is likely if the quantum computing cloud provider is rogue, (b) does not manipulate the quantum circuit. This is possible since tampering the quantum circuit may drastically alter the computation outcome which can be suspected,  (c) has the computational resources to analyze the program results to determine which qubit lines to tamper with, and (d) methods to mask the tampering from showing up as a significant change in errors (one such method is shown later in the paper using Example1).

\subsection{Attack Scenario}

The adversary in the proposed attack model (Fig. \ref{3}d) takes the form of a less reliable/untrusted quantum service provider while posing as a reliable/trusted hardware provider. The adversary then modifies the solution before reporting it to the users and seeks to minimally tamper with the output of the program, either by making the sub-optimal solution the optimal for the users or by lowering the probability of the most likely solution. For example, assuming a 3 qubit ($q_2,q_1,q_0$) quantum program that has optimal output of `100' while the next probable output being `101'. The adversary can target the $q_0$ and tamper the results such that `101' becomes the optimal output being sent to the user instead of `100'. This tampering can be accomplished in a variety of ways, one of which is to introduce a targeted bit flip error on the $q_0$ qubit line during measurement operation. Note that adversary has access to computation results (e.g., basis state probabilities of qubits) before sending to the user. Assuming that the quantum circuit itself is correct and optimal, the solution obtained by the adversary will be optimal which in turn can be tampered. In the following section, we will discuss some of the tampering models.

\subsection{Adversarial Tampering Model}

Various rogue providers may adopt their favorite method for tampering the results. Some examples are as follows.

\subsubsection{Random tampering}

While measuring the qubit lines, the adversary can introduce random bit flip error. In this type of tampering which has no overhead, the adversary tampers with the results by lowering the probability of the most likely solution. The adversary can introduce tampering in the form of qubit measurement error (Fig. \ref{4}), either on all of the qubit lines or on a subset of the qubit lines randomly.

\subsubsection{Targeted tampering}

In the case of targeted tampering, the attack will be more strategic in nature (Fig. \ref{4}) focusing on specific qubit lines to introduce measurement errors. The proposed algorithm for targeted tampering is described in Algorithm \ref{1}.

\begin{algorithm}
\SetAlgoLined
\KwIn{original bit-string counts}
\KwOut{tampered bit-string counts}
 Array a = most sampled correct bit string\\
 Array b = most sampled wrong bit string\\
\For{i $\in$ number of qubits in a bit string}{
    \uIf{$a_i$ $\neq$ $b_i$ }{
        add tampering to qubit-line i\;
    }
    \Else{
        no tampering;
    }
}
\KwOut{$tampered~bit~string~counts $}
\caption{Adversarial targeted tampering}
\label{1}
\end{algorithm}

\begin{figure}
    \centering
    \includegraphics[width= 3.25in]{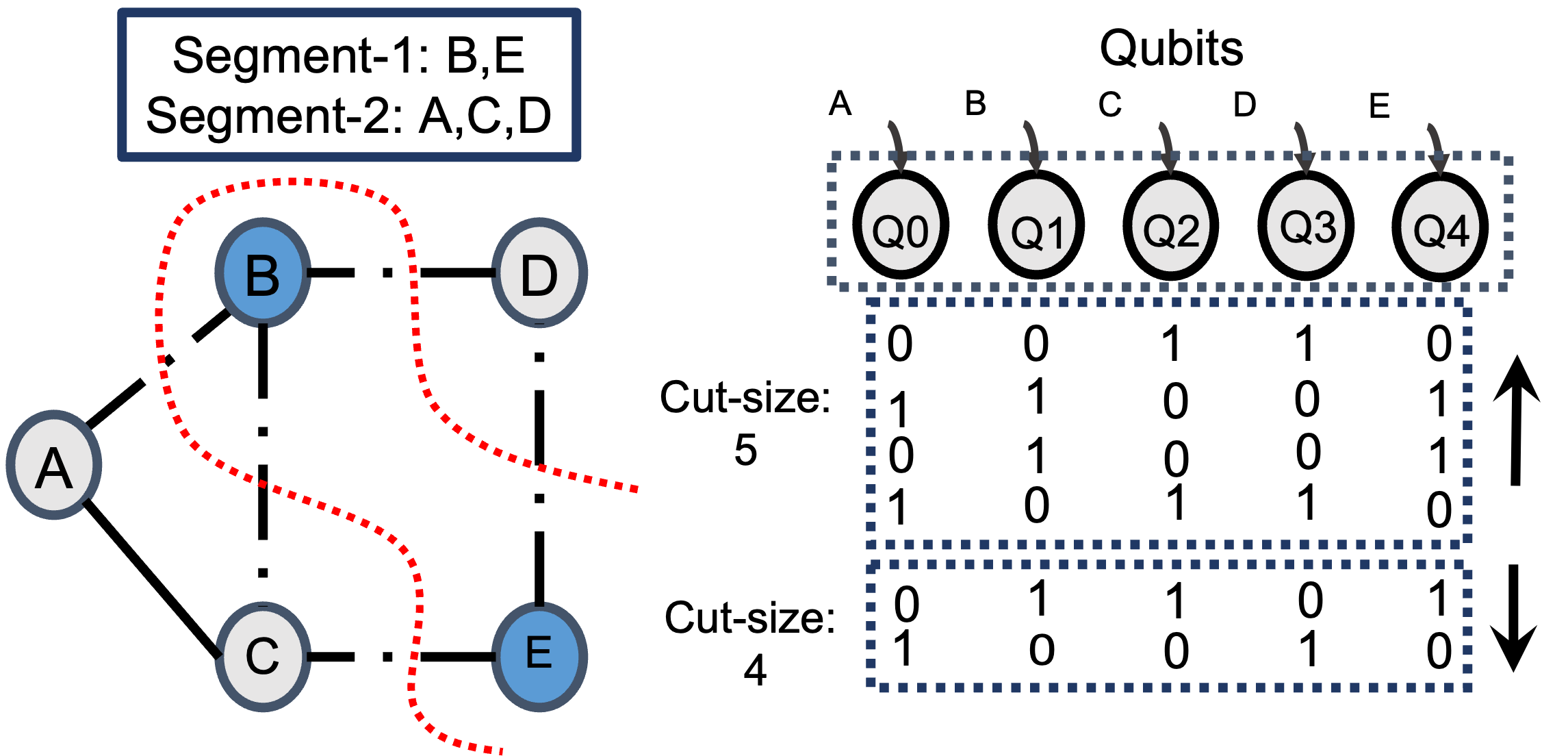}
    \caption{QAOA illustration for solving the maximum-cut (MaxCut) problem. The MaxCut size for the represented 5 node graph is 5 (cut in red). The probabilities of basis state measurements that represent larger cut-sizes for the problem graph are increased iteratively by QAOA. After the QAOA is completed, the states in cut size 5 will have higher probabilities than the states in cut size 4.
    }
    \label{15}
\vspace{-4mm}
\end{figure}

\section{Proposed Tampering model}
\subsection{Tampering Framework used for Simulations}

We model adversarial tampering by introducing extra measurement error on the qubit lines i.e., while performing final measurement on a qubit, we flip the state of the qubit with probability $t$, which we refer to in the paper as the \emph{tampering coefficient}. The proposed sample attack model is depicted in Fig. \ref{3}(d). We use IBM's fake backends to mimic real hardware and add this bit flip error as an attempt by the adversary to tamper with it. This bit flip error can be added to either a targeted qubit (per the Algorithm 1), to all qubit lines, or randomly selected qubit lines. The extra measurement error can be easily hidden when reported alongside readout errors. Consider following example to understand how an adversary can conceal the tampering while reporting measurement errors of the hardware. 


Example 1 : We consider the real hardware measurement errors, quoted as Readout assignment error (RAE), for the IBM's 5 qubit devices (Fig. \ref{3}(a)). Assuming tampering and RAE to be independent and uncorrelated sources of error, we can get the total error as:

\begin{equation}\label{Equation 2}
    \Delta RAE_{net}  = \sqrt{(\Delta RAE_{qi})^2 + (\Delta Tampering)^2 }
\end{equation}
where $(\Delta RAE_{qi}) = RAE$ value for $i^{th}$ qubit line and $\Delta Tampering$ is defined as :

\begin{equation}\label{eq:AR}
    \Delta Tampering  = t/n 
\end{equation}
where $t=$ tampering coefficient,
$n=$ (total qubit lines$ - $tampered qubit lines + 1). 

\begin{table}[]
    \centering
    \caption{Benchmark Characteristics}
    \begin{tabular}{cccccccc}
    \hline
     Benchmark  & Algorithm & Qubits & Gates\\ 
     \hline
     grover$_-$n2   & Search and Optimization  & 2 & 16 \\
     grover$_-$n3   & Search and Optimization  & 3 & 24 \\
     fredkin$_-$n3  & Logical Operation & 3 & 19 \\
     toffoli$_-$n3  & Logical Operation  & 3 & 18 \\
     hs4$_-$n4 & Hidden Subgroup  & 4 & 28 \\
    
     adder$_-$n4 & Quantum Arithmetic  & 4 & 23 \\
     
     inverseqft$_-$n4 & Hidden Subgroup  & 4 & 8 \\
     
     adder$_-$n10 & Quantum Arithmetic  & 10 & 142 \\
     
     multiply$_-$n13 & Quantum Arithmetic  & 13 & 98 \\
     \hline
     \end{tabular}
    \label{tab:1}
\end{table}

Assuming that the adversary uses ($t =0.1$ or $t =0.5$) , for ibmq$_-$lima $q_1$ RAE, we can calculate the final measurement error (which is 0.028 and 0.09, respectively) for that qubit line using \emph{equation} 2 for targeted tampering. 
These final error values are comparable to the values quoted for various devices Fig. \ref{3}a and qubit lines. For example the new RAE ($t =0.1$) for ibmq$_-$lima $q_1$ is comparable to RAE's of $q_0$, $q_2$ and even less than $q_4$.  
When the RAE value for $t =0.5$ is compared with the tamper-free RAE values of other hardware such as ibmq$_-$quito $q_3$, it is still found to be less.
However, our simulations show that an adversarial attack with a tampering coefficient ($t=0.1$). As a result, the adversary can easily get away with the tampering. 


\begin{figure}
    \centering
    \includegraphics[width= 3.25in]{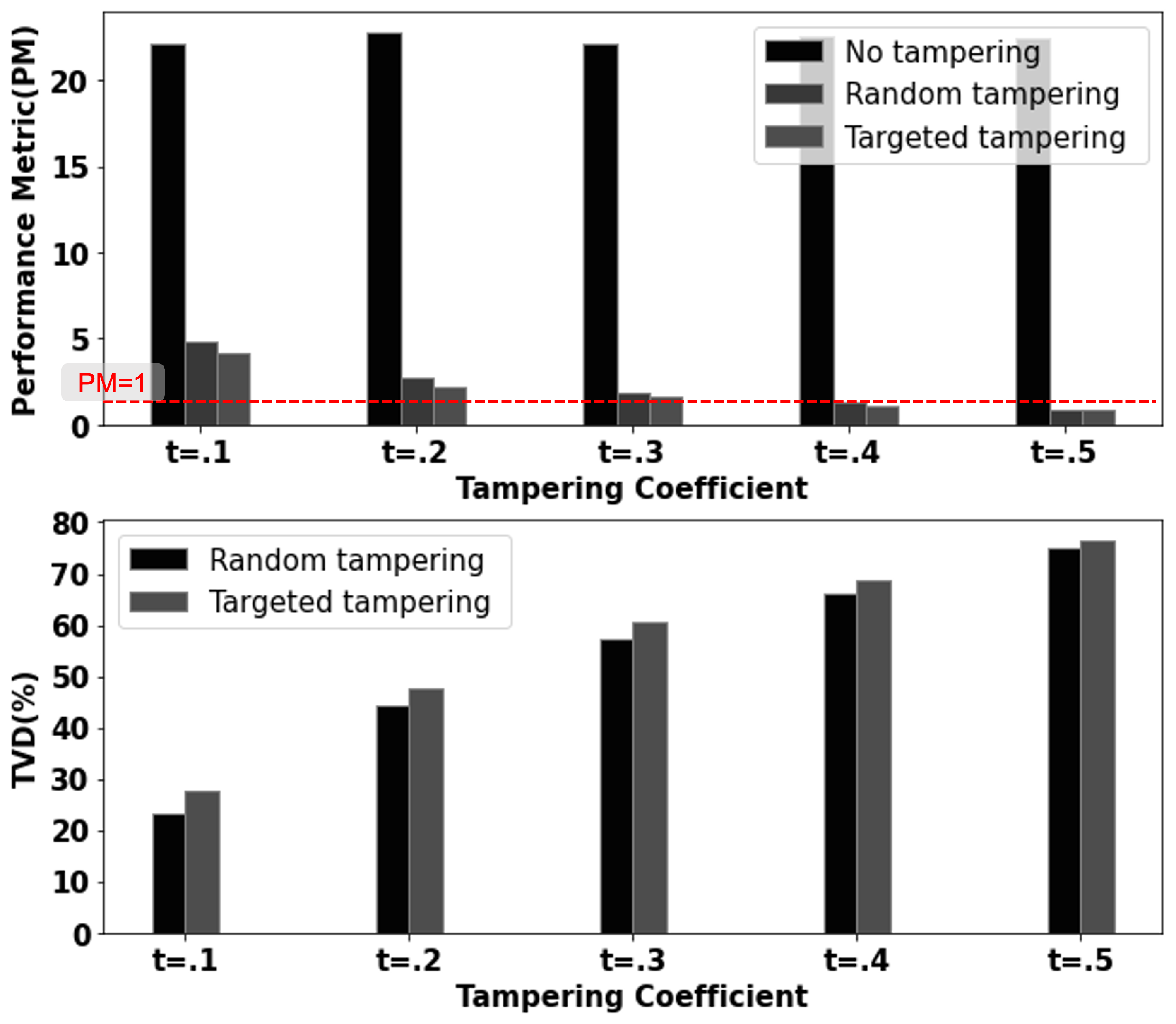}
    \caption{Comparison of PM and TVD for random vs targeted tampering. Benchmark (toffoli$_-$n3) is run for 10,000 shots on Fake$_-$montreal. Correct o/p: 111 (q2,q1,q0) and erroneous o/p with highest freq.: 101. Hence as per Algorithm 1 we introduce targeted bit flip error on q1. Random tampering is simulated by introducing bit flip error to all the qubit lines.}
    \label{5}
\vspace{-4mm}
\end{figure}

\begin{figure}
    \centering
    \includegraphics[width= 3.25in]{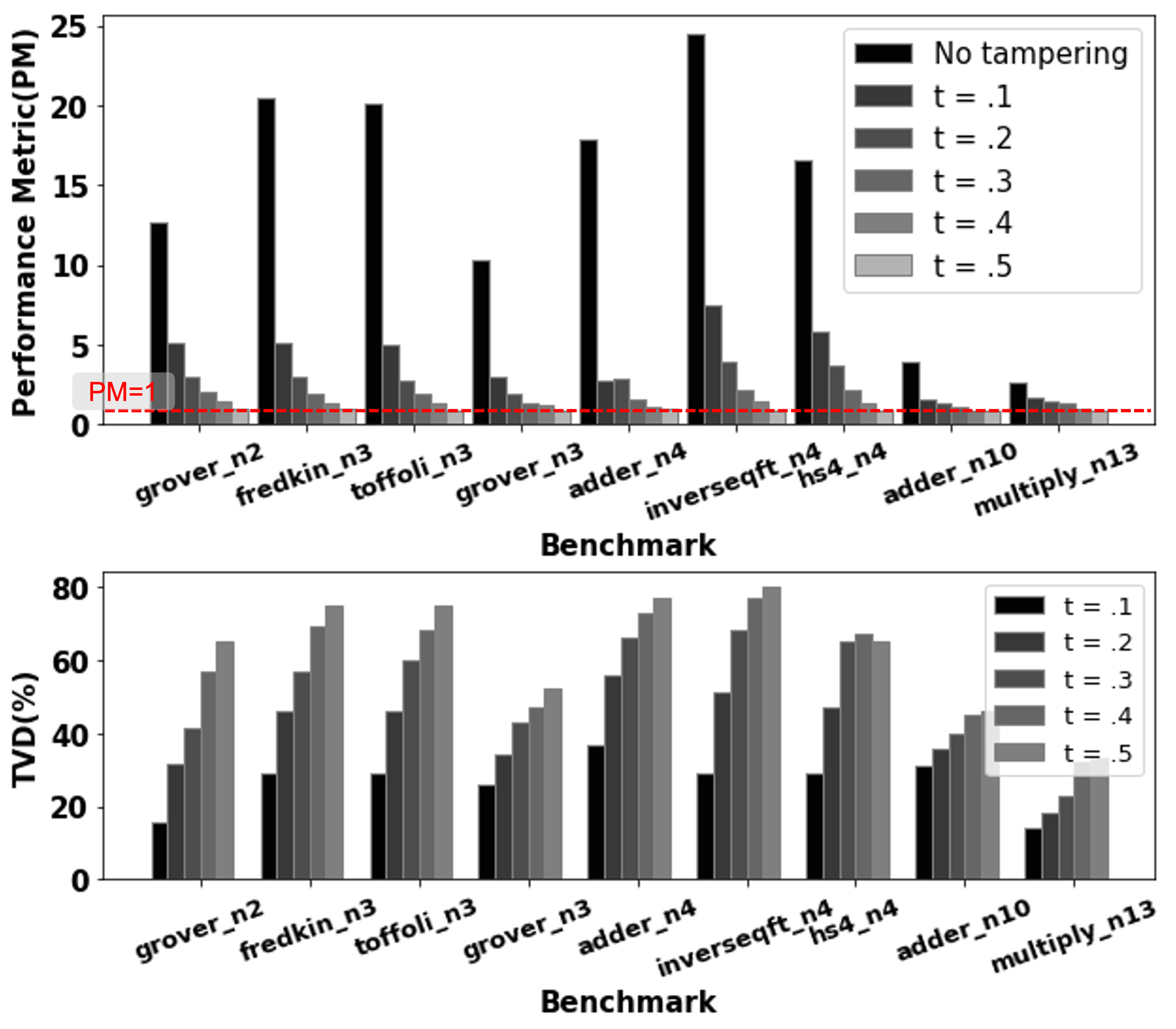}
    \caption{The PM and TVD variation with the tampering coefficient for various benchmarks. Each benchmark is run for 10,000 shots with varying degrees of adversarial tampering on the Fake$_-$montreal backend. A $PM < 1$ indicates that the hardware has converged to an incorrect result.
    }
    \label{6}
\vspace{-4mm}
\end{figure}

\subsection{Benchmarks}
We use the open-source quantum software development kit from IBM (Qiskit) \cite{b26} for simulations. A Python-based wrapper is built around Qiskit to accommodate the proposed attack model. Our benchmark suite includes 10 quantum benchmarks with 9 standard (i.e., pure quantum) circuits and 1 iterative (i.e., hybrid classical-quantum) algorithm i.e., QAOA as described below. 

\subsubsection{Pure quantum workloads}
The benchmark circuits/workloads include 2- and 3- qubit Grover search (grover$_-$n2 and grover$_-$n3), 3-qubit Fredkin gate (fredkin$_-$n3), 3-qubit Toffoli gate(toffoli$_-$n3), 4-qubit Hidden-Subgroup algorithm (hs4$_-$n4), 4 and 10-qubit Adder (adder$_-$n4 and adder$_-$n10), 4-qubit Inverse QFT (inverseqft$_-$n4), 13-qubit Multiply (multiply$_-$n13). These are adopted from QASMBench \cite{b27} which contains a low-level, benchmark suite based on the OpenQASM assembly representation. Selected benchmark suite covers a wide range of communication patterns, number of qubits, number of gates and depths that are needed to evaluate the proposed attack and defenses. Table \ref{tab:1} provides a summary of the chosen benchmarks.

\subsubsection{Hybrid quantum classical workload}

We use the iterative algorithm QAOA \cite{b14} to solve a combinatorial optimization problem MaxCut \cite{b24} to investigate the effects of adversarial tampering and its impact on the hybrid quantum-classical algorithms.
The MaxCut problem involves identification of a subset S$\in$V such that the number of edges between S and it's complementary subset is maximized for a given graph $G = (V, E) $with nodes V and edges E. MaxCut is NP-hard problem \cite{b24}, but there are effective polynomial time classical algorithms that can approximate the solution within a defined multiplicative factor of the optimum \cite{b25}. Using a p-level QAOA, an N-qubit quantum system is evolved with $H_-C$ and $H_-B$ p-times to find a MaxCut solution of an N-node graph Fig. \ref{16}. QAOA-MaxCut iteratively increases the probabilities of basis state measurements that represent larger cut-size for the problem graph. Qubits measured as 0’s and 1’s are in two different segments of the cut Fig. \ref{15}.

\subsection{Simulators and Hardware}

We use the fake provider module in Qiskit as noisy simulators to run our benchmarks. The fake provider module contains providers and backend classes. The fake backends are created using system snapshots to mimic the IBM Quantum systems. Important details about the quantum system, including coupling map, basis gates, and qubit parameters (T1, T2, error rate, etc.), are contained in the system snapshots. We use following backends \emph{(mimicking their actual hardware representations)} for our experiments:  Fake$_-$Vigo (5 qubit), Fake$_-$Athens (5 qubit), Fake$_-$Yorktown (5 qubit), Fake$_-$Melbourne (14 qubit), Fake$_-$Tokyo (20 qubit), Fake$_-$Montreal (27 qubit), Fake$_-$Mumbai (27 qubit), Fake$_-$Paris (27 qubit), Fake$_-$Toronto (27 qubit), Fake$_-$Manhattan (65 qubit). We also run some of our benchmarks on ibmq$_-$manila (actual quantum hardware provided by IBM) and use the modeled tampering parameters for proof-of-concept demonstration of the proposed tampering and defenses.

\subsection{Evaluation Metric}

\subsubsection{Performance Metric (PM)}

We define Performance Metric (PM) (which is the ratio of the probability of correct and the most frequent incorrect basis states, Fig. \ref{3}b) as a way to measure how well a NISQ machine can infer the right response. PM greater that 1 indicates that the system will be able to correctly infer the output. For PM less than 1, the wrong answer(s) would mask out the correct answer. As our objective is quantify the effect of tampering on sampled output, we use PM as the primary figure of merit in evaluations. This metric has also been used in previous works for performance quantification \cite{b28}. 

\subsubsection{Total Variation Distance (TVD)}

We also use total variation distance (TVD) (Fig. \ref{3}c) as an additional metric to quantify the impact of tampering on the probability distribution of the output states for a given program. The quantum circuit outcome is a distribution of probabilities or ‘counts’ of possible binary states. TVD compares the probabilities of the same binary states between two distributions. If the probabilities are identical, then TVD = 0. The TVD will also be higher for extremely diverse distributions.

\begin{figure}
    \centering
    \includegraphics[width= 3.25in]{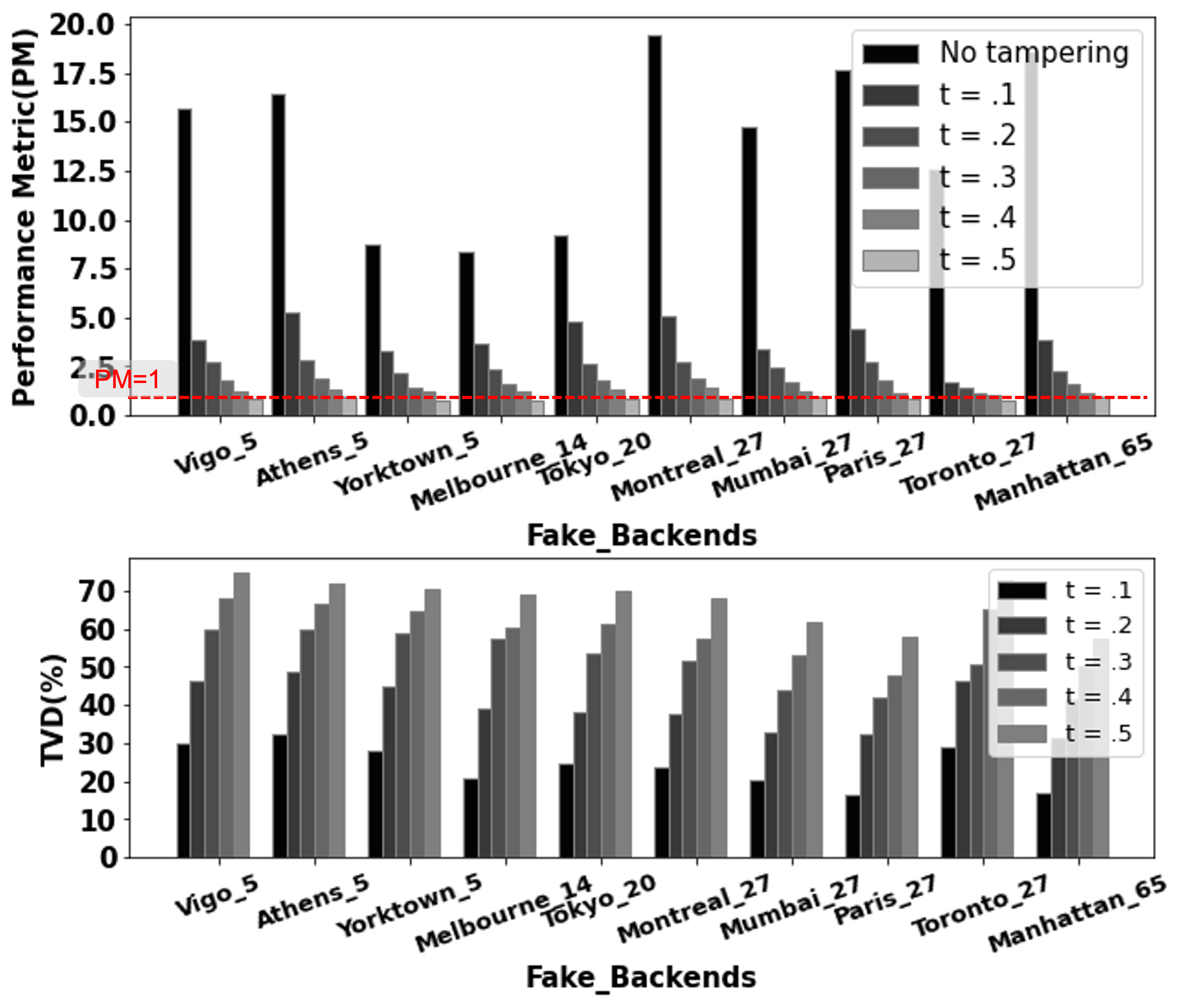}
    \caption{PM and TVD variation with the tampering coefficient for benchmark toffoli$_-$n3 across various fake back-ends. All tested fake backends fail to converge to the correct solution at t =0.5.
    }
    \label{7}
\vspace{-4mm}
\end{figure}

\subsection{Simulation and Results}


\subsubsection{Random vs targeted tampering}

The performance of a randomly tampered version and a targeted tampered version of the Fake$_-$montreal backend is evaluated. The Fig. \ref{5} depicts the PM and TVD variation of the benchmark toffoli$_-$n3 after 10,000 shots on these tempered back-ends. We introduce bit flip error on all qubit lines during measurement to account for random tampering. However, for targeted tampering, we choose q$_1$ (as per the Algorithm1) to introduce the bit flip error. 
We note a 75$\%$ reduction in PM for random tampering and an 80$\%$ reduction in PM for minimal tampering (t =0.1). Therefore, the attack is able to degrade the resilience of computation significantly. At t=0.5, the PM for all programs becomes less than 1, indicating that the correct result cannot be inferred. For t =0.1, the probability distributions for random and targeted tampering differ by 24$\%$ and 29$\%$, respectively, with a very high TVD of 70$\%$ and 72$\%$ for random and targeted tampering, respectively, when t =0.5. We use targeted tampering in the simulations that follow throughout the paper since it is more effective in degrading performance.

\begin{table}[]
    \centering
    \caption{PM vs Shots (tampering coefficient =0.5) ($_-$t denotes tampered results)}
    \begin{tabular}{cccccccc}
    \hline
      &  & & No. of shots &  &\\ 
     \hline
     Benchmark & 500 & 1000 & 2000 & 5000 & 10000\\ 
     \hline
    $grover_-2$ & 11 & 13.5 & 11 & 11.8 & 12.4\\
    $grover_-2_-t$ & .92 & .93 & .91 & .90 & .87\\
     \hline
    $fredkin_-3$ & 23.6 & 24.3 & 26.5 & 31.7 & 26.3\\
    $fredkin_-3_-t$ & .82 & .79 & .78 & .89 & .77\\
     \hline
    $toffoli_-3$ & 19.2 & 18.2 & 21.4 & 22.2 & 23.3\\
    $toffoli_-3_-t$ & .89 & .90 & .85 & .83 & .92\\
     \hline
    $grover_-3$ & 9.3 & 8.9 & 10 & 9.3 & 9.3\\
    $grover_-3_-t$ & .87 & .74 & .96 & .92 & .84\\
     \hline
    $adder_-4$ & 18.2 & 19.1 & 18.9 & 19 & 17.3\\
    $adder_-4_-t$ & .85 & .93 & .89 & .86 & .95\\
      \hline
    $inverseqft_-4$ & 20.4 & 23.5 & 25.1 & 20.2 & 21.4\\
    $inverseqft_-4_-t$ & .72 & .70 & .74 & .87 & .94\\
      \hline
    $hs4_-4$ & 18.5 & 19.4 & 18.34& 17.5 & 16.6\\
    $hs4_-4_-t$ & .72 & .68 & .96 & .95 & .96\\
      \hline
    $adder_-10$ & 3.2 & 4.3 & 3.4 & 3.3 & 2.4\\
    $adder_-10_-t$ & .64 & .76 & .74 & .91 & .91\\
      \hline
    $multiply_-13$ & 2.4 & 2.0 & 1.9 & 2.2 & 2.4\\
    $multiply_-13_-t$ & .71 & .85 & .79 & .80 & .87\\
     \hline
     \end{tabular}
    \label{tab:2}
\end{table}

\begin{figure}
    \centering
    \includegraphics[width= 3.25in]{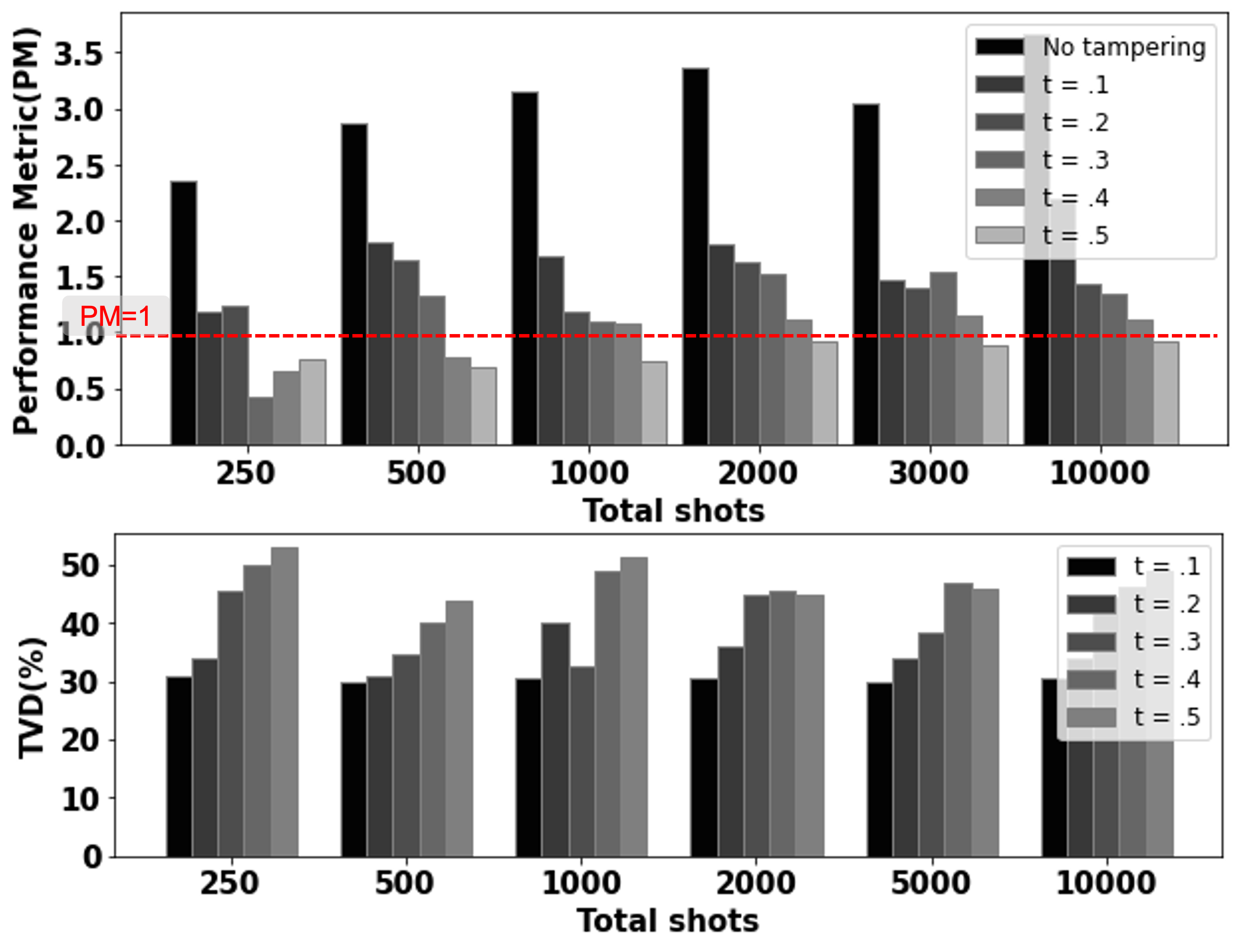}
    \caption{Adder$_-$n10 simulated for a variable number of shots on backend fake$_-$montreal. Adversarial tampering is more effective if the user has less number of available shots. For the given benchmark, the PM degrades below 1 for just t=0.3 when run for 250 shots.
    }
    \label{8}
\vspace{-4mm}
\end{figure}

 \begin{figure}
    \centering
    \includegraphics[width= 3.25in]{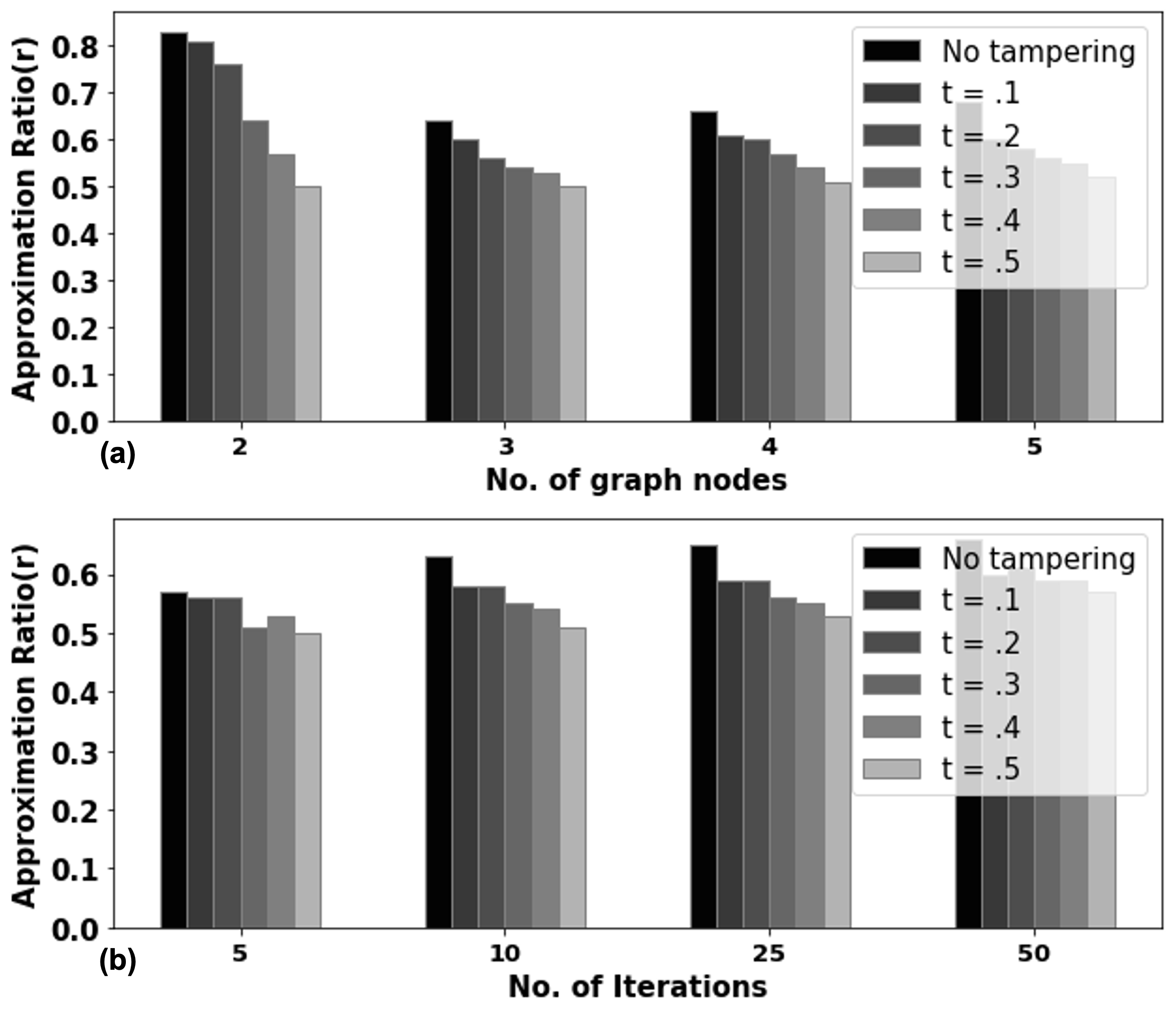}
    \caption{a) Approximation ratio (r) variation for different graph sizes when run on tampered hardware fake$_-$montreal for 50 iterations. b) Performance comparison of QAOA for four-node graph with varying iterations on fake$_-$montreal. }
    \label{17}
\vspace{-4mm}
\end{figure}

\subsubsection{Impact of adversarial tampering on hardware performance}

Fig. \ref{6} depicts the performance metric (PM) and TVD variation with the tampering coefficient (which quantifies the amount of adversarial tampering) for the various programs in our benchmark suite. Each benchmark is run on the Fake$_-$montreal backend for 10,000 shots. A  $PM < 1$ indicates that the hardware converges to the wrong result, i.e., the probability of the correct solution falls below the probability of the other incorrect states. PM should ideally be as high as possible (at least greater than 1.0). A high PM ($PM > 2.5$) indicates a dominant correct solution with a high probability of being correct. On the contrary, a lower value is preferred for TVD metric. TVD denotes a difference in the probability distribution between tamper-free and tampered hardware. As a result, the adversarial attack seeks to reduce the PM and increase the TVD for the programs, degrading the overall computation performance. The simulation result Fig. \ref{6} shows that as tampering coefficient is increased, the PM for all benchmarks drops significantly ($\approx$ 65$\%$ on average at just t =0.1) and a high TVD is observed. This pattern holds true across all benchmarks. Furthermore, at t =0.5, the PM for all programs falls below 1, indicating that the correct result from cannot be inferred with reasonable confidence.

Following that, we run the benchmark (toffoli$_-$n3) for 10,000 shots across 10 fake backends to quantify the effect of the proposed tampering model on different hardware with varying number of available qubits, qubit connectivity, error rates, and so on. The PM and TVD variations with tampering 
is depicted in Fig. \ref{7}. The same trend of PM degradation ($\approx$ 68$\%$ on average at just t =0.1) and significant increase in TVD is observed. For t =0.5, all of the tested fake backends fail to converge to the correct solution.

\subsubsection{Effect of tampering with varying number of shots}

 \begin{figure}
    \centering
    \includegraphics[width= 3.25in]{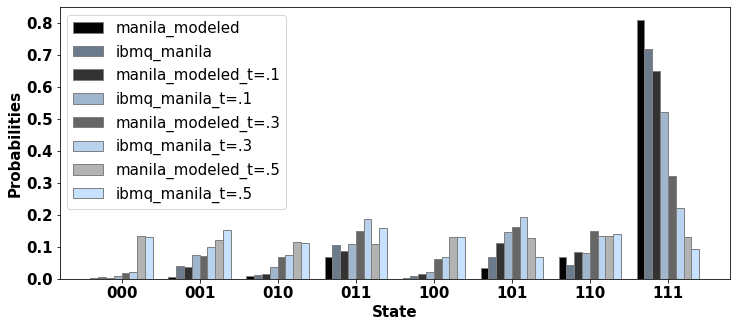}
    \caption{The proposed tampering model is used to compare the performance of a fake backend (manila$_-$modeled) modeled from the real hardware (ibmq$_-$ manila) and the actual hardware (ibmq$_-$manila). We observe that the tampering results are comparable, with only minor variations which can be attributed to temporal variations.
    }
    \label{13}
\end{figure}

\begin{figure}
    \centering
    \includegraphics[width= 3.25in]{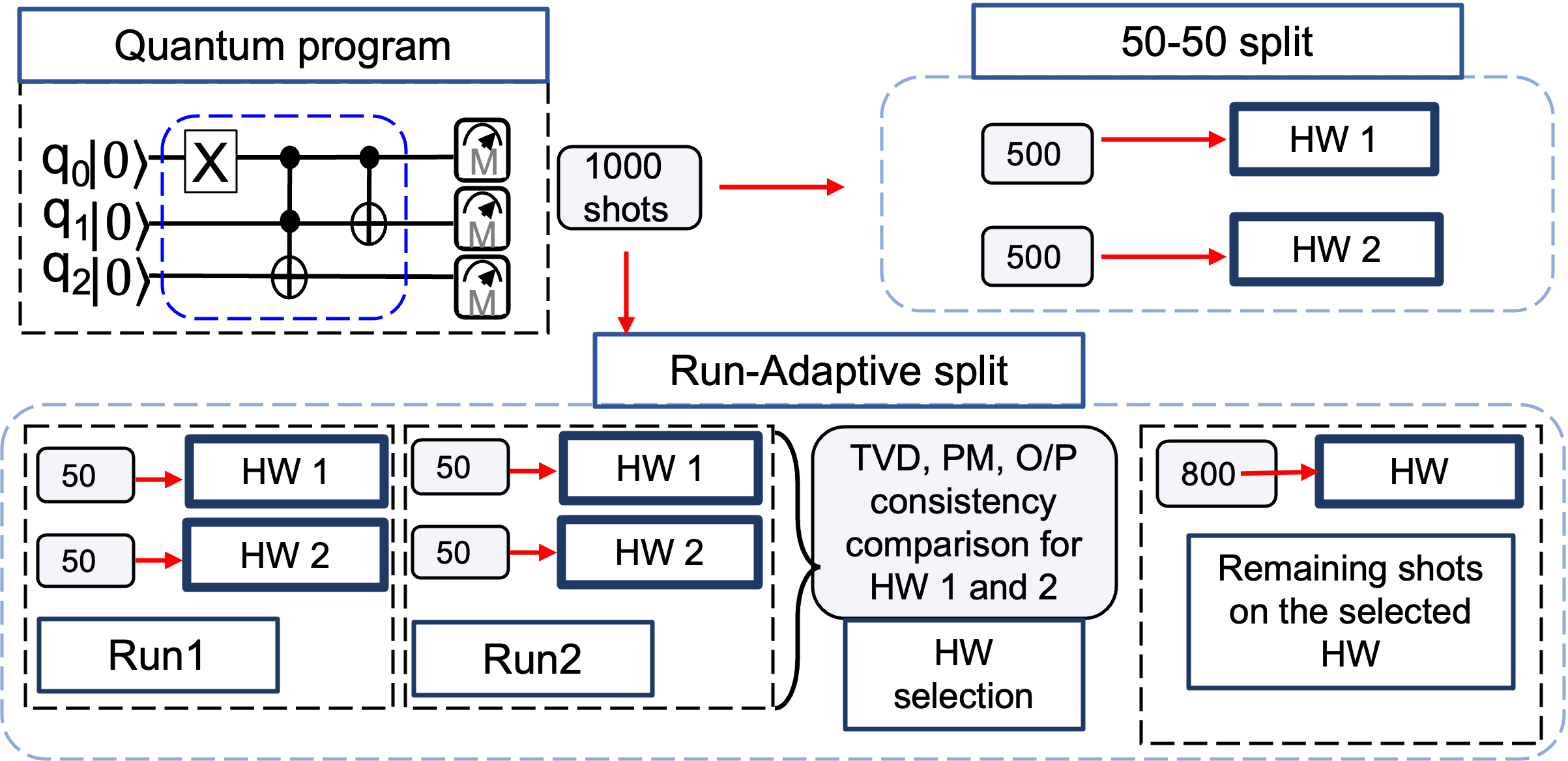}
    \caption{Two different shot splitting approaches to mitigate the effect of adversarial tampering. The user can either do a 50-50 split, where the shots are distributed equally on available hardware and the results are stitched together to get the converged answer, or the user can start with two initial runs of small number of shots (say, 50) on both hardware, compare PM, TVD, and output confidence, and run the rest of the remaining shots on the hardware that appears to be better. }
    \label{9}
\vspace{-4mm}
\end{figure}

On fake$_-$montreal, we run various benchmarks with different number of qubits, depth, and gate sizes by varying the number of shots for t =0.5. The results are summarized in the Table \ref{tab:2}. Even 10,000 shots for a 2-qubit program (grover$_-$2) is insufficient to achieve correct convergence. Fig. \ref{8} depicts a 10-qubit program (adder$_-$n10) simulated for a variable number of shots. When programs with a large number of qubits are run for a small number of shots, adversarial tampering has a greater impact. In the case of benchmark adder$_-$n10, the PM degrades below 1 for t=0.3 when run for 250 shots, and we see a similar degradation at much stronger tampering t =0.5 when run for 1000 shots and above. When run for 250 shots, the average TVD increase is also maximum ($\approx$ 50\%).

\begin{table*}[]
    \centering
    \caption{Intelligent shot distribution : Identifying tampered/bad hardware (shots/run = 50 ; $_-$t denotes tampered results)}
    \begin{tabular}{ccccccc||ccccc}
    \hline
      &  & & & & $toffoli_-n3$ & & & & & $adder_-n10$ &\\ 
     \hline
      & & & PM & TVD(\%) & Most Freq. O/P & Probability &  & PM &  TVD(\%)   & Most Freq. O/P & Probability \\ 
     \hline
    $t=0.1$ &  $HW$ & Run 1 & 22 &  & 111 & 44/50 & & 4.75 &  & 10000 & 19/50\\
           &          & Run 2 & 11 & 7 & 111 & 44/50 & & 2.71 & .9 & 10000 & 19/50\\
           & $HW_-t$  & Run 1 & 2.77 &  & 111 & 24/50 & & .8 &  & 10111 & 5/50\\
           &          & Run 2 & 2.66 & 13 & 111 & 25/50 & & .5 & 18 & 00000 & 4/50\\
     \hline
     
    $t=0.2$ &  $HW$ & Run 1 & 22 &  & 111 & 44/50 & & 2 &  & 10000 & 16/50\\
           &          & Run 2 & 8.4 & 6 & 111 & 42/50 & & 3.2 & 1.2 & 10000 & 16/50\\
           & $HW_-t$  & Run 1 & 2.22 &  & 111 & 20/50 & & .2 &  & 01010 & 4/50\\
           &          & Run 2 & 1.72 & 12 & 111 & 19/50 & & .25 & 15 & 00000,01110 & 4/50\\
     \hline
    $t=0.3$ &  $HW$ &Run 1 & 14.33 & & 111 & 43/50 & & 4.2 &  & 10000 & 21/50\\
           &          & Run 2 & 21 & 6 & 111 & 45/50 & & 3 & 1 & 10000 & 21/50\\
           & $HW_-t$  & Run 1 & 1.53 &  & 111 & 16/50 & & .43 &  & 01100 & 4/50\\
           &          & Run 2 & 1.6 & 15 & 111 & 17/50 & & .5 & 17 & 00000 & 4/50\\
     \hline
    $t=0.4$ &  $HW$ & Run 1 & 14.33 &  & 111 & 43/50 & & 2.5 &  & 10000 & 18/50\\
           &          & Run 2 & 10.74 & 6 & 111 & 43/50 & & 4.5 & 1.2 & 10000 & 18/50\\
           & $HW_-t$  & Run 1 & .91 &  & 101 & 11/50 & & .6 &  & 00011 & 5/50\\
           &          & Run 2 & 1.37 & 17 & 111 & 11/50 & & .5 & 19  & 00001 & 3/50\\   
     \hline
    
    $t=0.5$ &  $HW$ & Run 1 & 23.5 &  & 111 & 47/50 & & 2 &  & 10000 & 16/50\\
           &          & Run 2 & 20.5 & 1 & 111 & 41/50 & & 3.6 & 1 & 10000 & 18/50\\
           & $HW_-t$  & Run 1 & .62 &  & 010,000 & 8/50 & & .5 &  & 10110 & 4/50\\
           &          & Run 2 & .66 & 10 & 101 & 9/50 & & .28 & 14 & 11110 & 7/50\\
     \hline
     \end{tabular}
    \label{tab:3}
\end{table*}

\subsubsection{Impact of tampering on QAOA performance}

For the sake of simplicity, we will focus on MaxCut on unweighted d-regular graphs (UdR), where each vertex is connected to only adjacent vertices. We use the approximation ratio defined in equation\ref{eq:AR} as the performance metric for QAOA. We run QAOA for each node graph ten different times and report the average values for the approximation ratio (r). The greater the r value, the better the performance. Ideally, the performance of QAOA can improve as p increases, with r $\rightarrow$ 1 when p $\rightarrow$ $\infty$. For our simulations, we run QAOA for maxcut on U2R, U3R, U4R, and U5R graphs to investigate the effects of adversarial tampering on quantum-classical hybrid algorithms.
Fig. \ref{17}a shows the variation in AR for QAOA solving maxcut for various graph nodes. In each case, we run QAOA (p=1) for 50 iterations (50 shots/iteration). We note $8\%$ and $25\%$ average reduction in AR for t=0.1 and t=0.5, respectively. Fig. \ref{17}b depicts the variation in AR with the number of iterations for a 4-node graph run on tampered hardware with varying degrees of tampering. When the number of iterations available to users for the tampered hardware is limited, we see a decrease in approximation ratio for a fixed tampering constant. When we run QAOA for 10 iterations rather than 50, AR degrades by 10$\%$ for t=0.1 and 25$\%$ for t=0.5, indicating that the performance of the hybrid-classical algorithm QAOA is sensitive to the number of available iterations when run on tampered hardware. However, reducing the number of iterations from 50 to 10 for tamper-free hardware results in a marginal ($2\%$) decrease in AR.

\subsection{Modeling Tampering on Real Hardware}

We created fake backends to simulate real hardware and test adversarial tampering and proposed solutions in our experiments. The majority of the hardware available to users is limited in terms of qubits and frequently has long queues for a single run. Fake backends are useful for testing a variety of benchmarks because they provide program performance comparable to real-time hardware. Since adding bit flip errors during measurement on real hardware is not possible, we model a back-end to mimic the real hardware using Qiskit's fake provider module. Then, for a specific program, we mimic the effects of tampering on the backend we built and use the results to model tampering on real hardware for that particular benchmark. The Fig. \ref{13} compares the performance of the fake backend (fake$_-$manila) that we modeled from the real IBM hardware (ibmq$_-$manila) and the real hardware (ibmq$_-$manila). We simulate adversarial tampering by running benchmark toffoli$_-n3$ for 10,000 shots on the simulator and actual hardware. We find that the probability distributions are very similar. The effect of modeled tampering is similar to the trends seen with the fake backend. As the extent of adversarial tampering (t) increases, the probability of the correct output `111' decreases in both cases, and at t=0.5, both the simulator and hardware fail to converge to the correct answer `111'.

\subsection{Summary of Tampering Analysis}
(a)  The adversary uses targeted tampering to degrade performance at the expense of computing the solution to the user's program from raw data. (b) Even minor tampering (t=0.1) is enough to reduce the confidence in the correct output across multiple benchmarks. (c) For benchmarks with a high qubit count, such as, adder$_-$n10 and multiply$_-$n13, minimal tampering is sufficient to change the output, resulting in users receiving a sub-optimal solution. (d) Users are more sensitive to tampering when fewer shots are used for a given program. (e) Tampering (t=0.1 to 0.5) can be easily masked by the adversary since the change in measurement error is negligible. (f) Performance of quantum-classical hybrid workloads is very sensitive to number of iterations when run on tampered hardware.

\begin{figure}
    \centering
    \includegraphics[width= 3.25in]{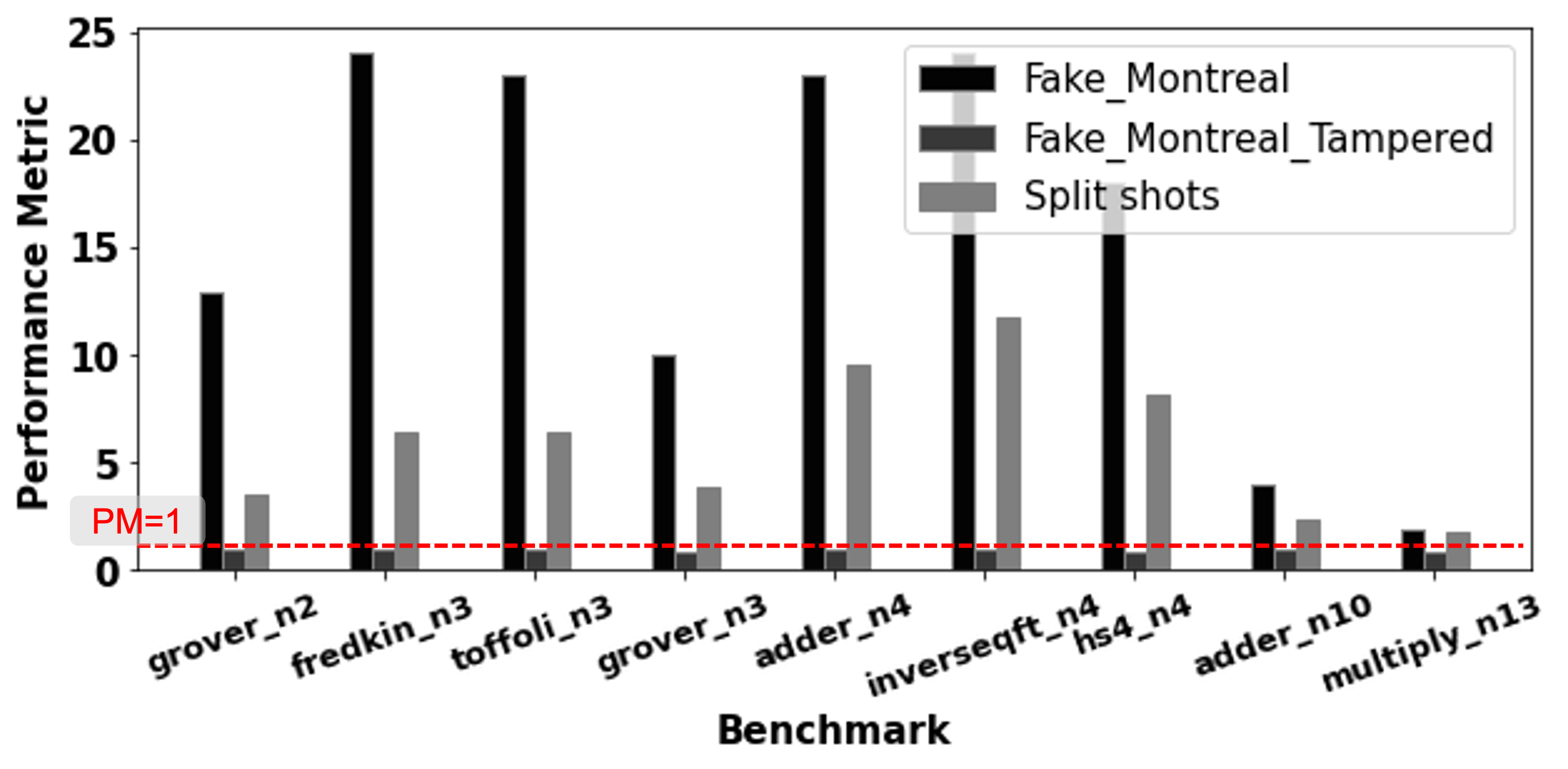}
    \caption{With t=0.5 and 10,000 shots, we simulate one tamper-free (Fake$ _-$montreal) and one tampered hardware (Fake$_-$montreal$_-$ tampered). A 50-50 split results in a significant improvement in PM ($\approx$300\% on average).
    }
    \label{10}
\vspace{-4mm}
\end{figure}

\section{Proposed Defenses}

\subsection{Basic Idea and Assumptions} 

We assume that, out of $n$ hardware options available to the user, at least one is reliable, i.e., tamper-free (we show results for up to 2 tampered hardware out of 3). Furthermore, user may avail the services of multiple untrusted cloud vendors which may have different tampering model. However, the user is unaware of tamper-free and tampered hardware and the adversarial tampering model.

\begin{figure}
    \centering
    \includegraphics[width= 3.25in]{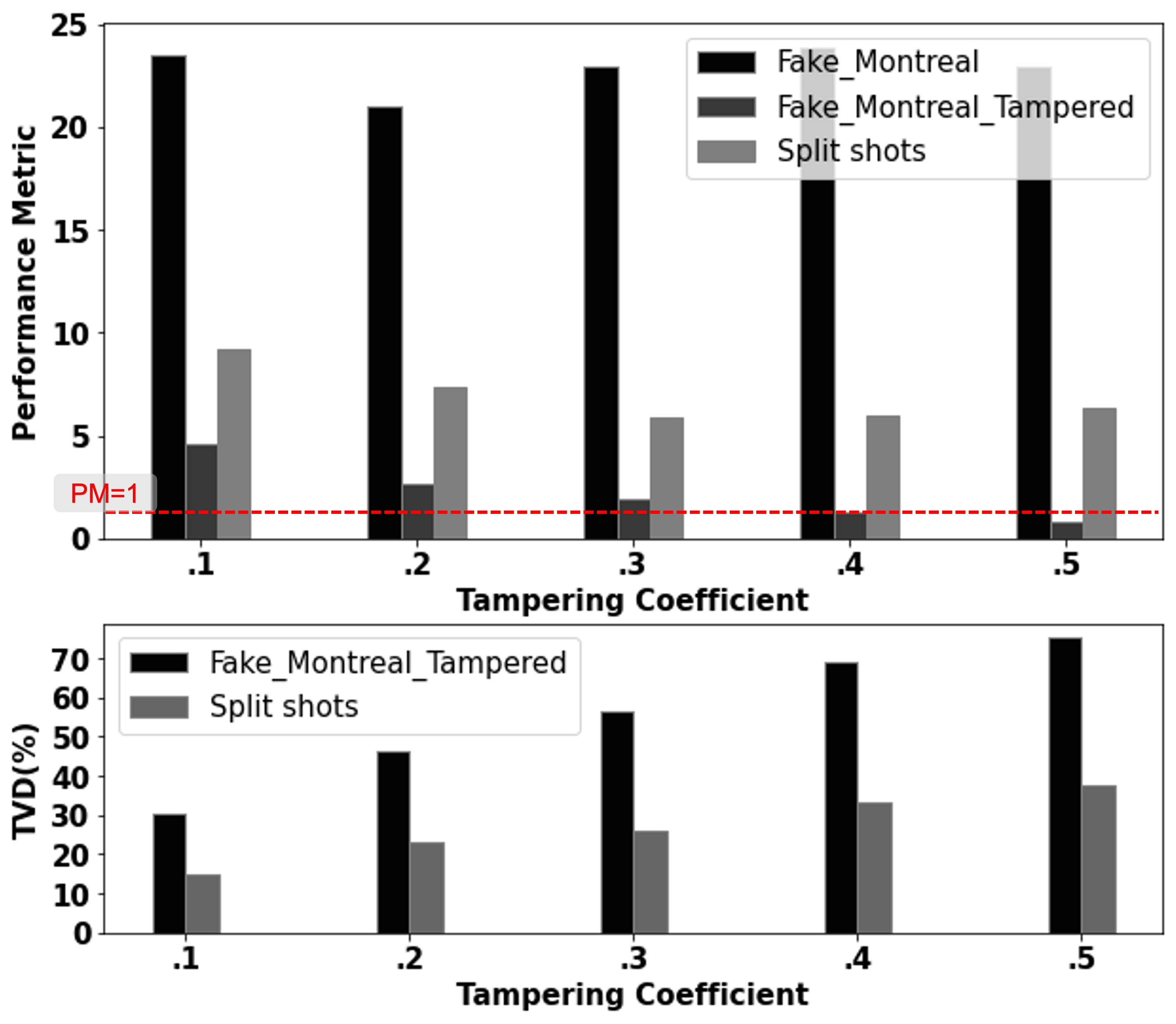}
    \caption{Improvement in PM and TVD for varying degrees of tampering. TVD reduction of $\approx$ 55$\%$ on average and improvement in PM margin ($\approx$ $125\%$ on average and a maximum of $\approx$ 400$\%$ for t=0.5) is observed for 50-50 split for toffoli$_-$n3 circuit.
    } 
    \label{11}
\vspace{-2mm}
\end{figure}

\begin{figure}
    \centering
    \includegraphics[width= 3.25in]{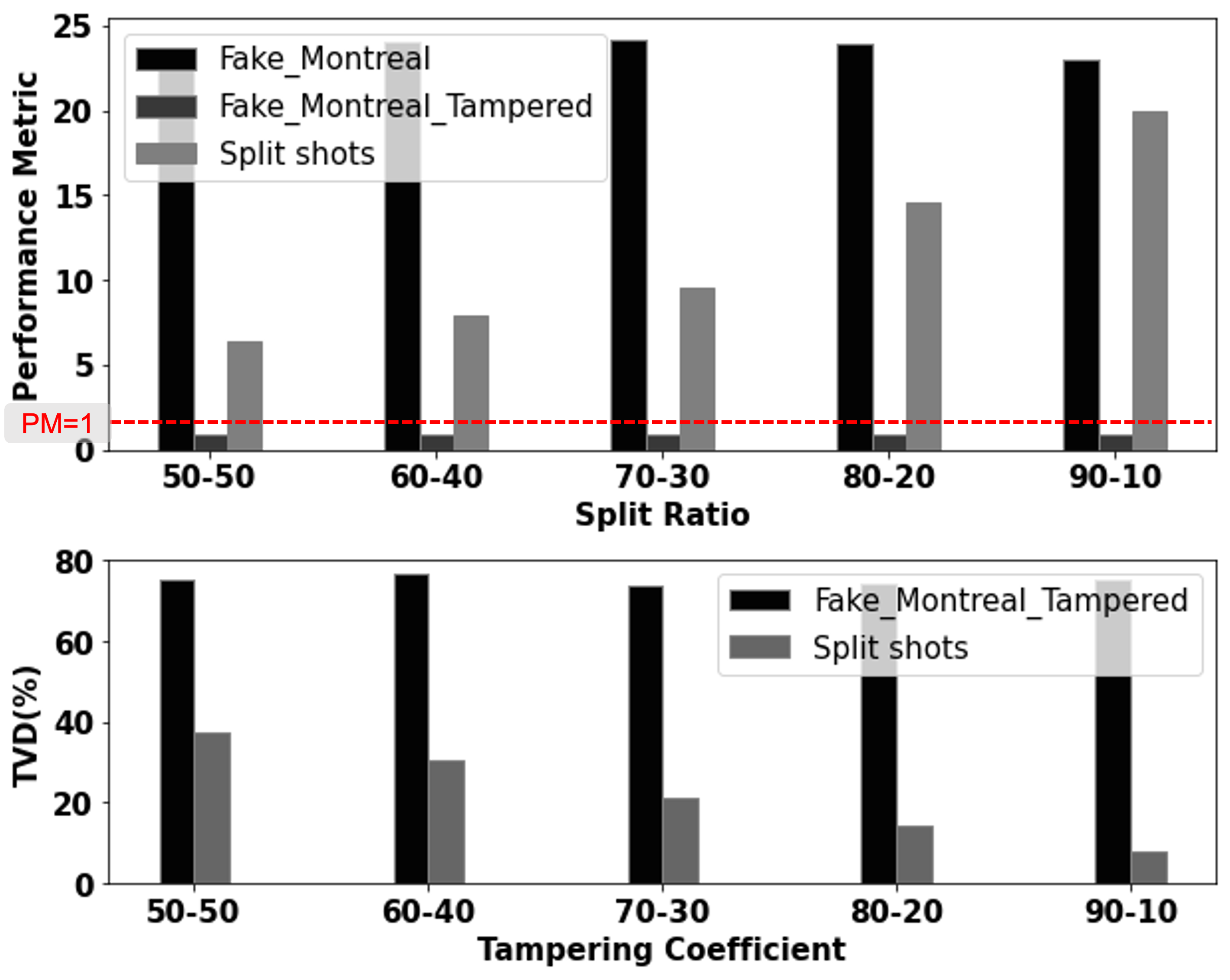}
    \caption{PM and TVD improvements as a function of the percentage of shots run on tamper-free hardware. The benchmark toffoli$_-$n3 is run on Fake$_-$montreal (tamper-free hardware) and Fake$_-$montreal$_-$tampered (tampered with t =0.5) for 10,000 shots. If the majority of shots are run on tamper-free hardware, the user can significantly improve resilience against tampering.
    }
    \label{12}
\vspace{-4mm}
\end{figure}

We propose splitting shots on available hardware (from different trusted and untrusted vendors) to mitigate the effects of adversarial tampering. For example, one may assume hardware from well-established AWS or IBM to be trusted and from less-established vendor X (may be located in an untrusted country) to be untrusted. Since untrusted/reliable hardware will provide correct solution, the chances of masking the tampered results is higher with shot splitting. Even if the vendors are untrusted, their tampering model may differ. Therefore, splitting the shots may increase the likelihood of suppressing the incorrect outcomes and obtaining correct outcome. We explain the methodology, PM, and reliability enhancements provided by the proposed shot distribution strategies below. The summary of the two shot splitting approaches is shown in Fig. \ref{9}.

\subsection{Equal Shot Distribution}

The user can divide the shots evenly among the available hardware without incurring any computational overhead (assuming the hardware are homogeneous and queuing delays are identical). For example, assuming the user has access to hardware HW1 and HW2 provided by two different service providers. HW2 however is plagued with tampering. User has to run a program P1 with 1000 shots. If he runs all those 1000 shots on HW2, the results received will be tampered and unreliable. Therefore, we split the shots between these hardware equally to make the results more resilient, thereby mitigating the adversary's tampering (Fig. \ref{9}). This can be generalized to a scenario with n number of available hardware.

\begin{figure}
    \centering
    \includegraphics[width= 3.25in]{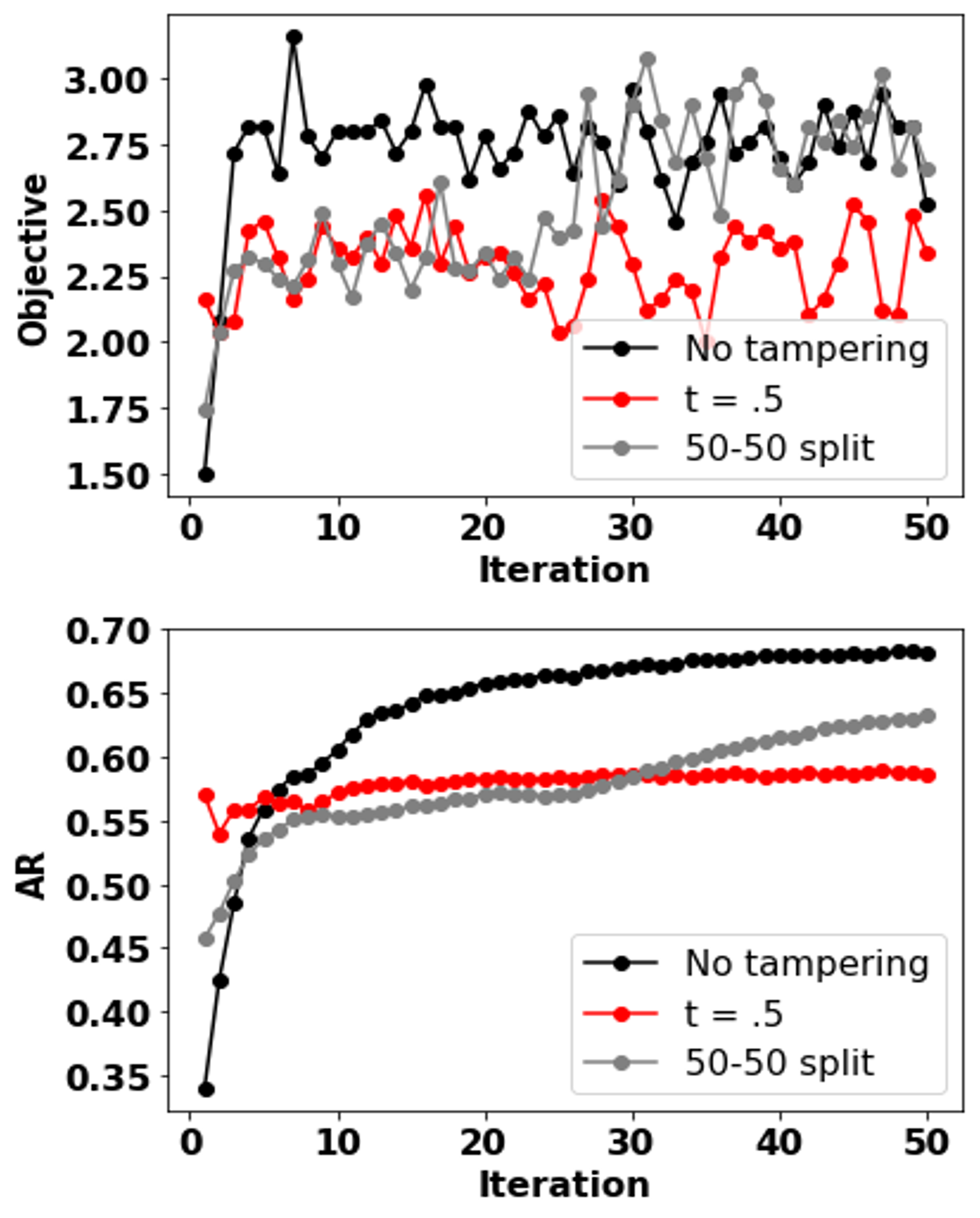}
    \caption{Effect of tampering (t=.5) on the objective and AR for a 4-node graph over 50 iterations and proposed 50-50 iteration split defense.}
    \label{18}
\end{figure}

\subsection{Adaptive Shot Distribution}

The user can also intelligently and adaptively distribute the shots. This can be done by running a few initial shots on all available hardware, comparing the results, doing majority voting, and then running the rest of the shots on the hardware that is more reliable. \emph{The characteristics of a dependable/tamper-free hardware is that it will converge to a specific solution for each batch of shots run on it, will have minimal increase in TVD between batches, and will have higher PM. The tampered hardware may return a different solution for each batch of shots run on it due to temporal variation in qubit quality and will exhibit higher TVD and lower PM among the batch of runs}. For example, we assume the case of HW1 and HW2 again, with user having 1000 shots to run. For Run 1, the user fires 50 shots on HW1 and HW2, records the results, and fires 50 shots again as Run 2. The TVD, PM, and repeatability of the final answer from the two iterations are then compared. The user will look for low TVD, high PM, and repeatable converged output and it's confidence (probability) to determine the best hardware (Fig. \ref{9}). In this way, the user can choose a better hardware to allocate the remaining shots. If not, another iteration of 50 shots on each hardware can be fired and the process is repeated until the user is satisfied. 
When more than two hardware is available, the same procedure can be used, and majority voting can be done to select the tamper-free hardware from a batch of given hardware.

\subsection{Simulation Results}


\begin{table}[]
    \centering
    \caption{AR vs Tampering (Iterations=50, Split=50:50)($_-$t denotes tampered results)}
    \begin{tabular}{cccccccc}
    \hline
      &  &  & Graph nodes &  &\\ 
     \hline
      &  & 2 & 3 & 4 & 5 \\ 
     \hline
    $t=0.1$ & HW & .84 & .63 & .68 & .68 \\
     & HW$_-$t & .78 & .6 & .65 & .65 \\
     & Split & .81 & .61 & .67 & .65 \\
     \hline
    $t=0.2$ & HW & .84 & .63 & .68 & .68  \\
     & HW$_-$t & .76 & .56 & .61 & .58 \\
     & Split & .78 & .61 & .65 & .62 \\
     \hline
    $t=0.3$ & HW & .84 & .63 & .68 & .68 \\
     & HW$_-$t & .67 & .54 & .57 & .56 \\
     & Split & .7 & .59 & .61 & .63 \\
     \hline
    $t=0.4$ & HW & .84 & .63 & .68 & .68 \\
     & HW$_-$t & .57 & .53 & .54 & .55 \\
     & Split & .65 & .6 & .58 & .63 \\
     \hline
    $t=0.5$ & HW & .84 & .63 & .68 & .68 \\
     & HW$_-$t & .5 & .5 & .5 & .55 \\
     & Split & .62 & .59 & .57 & .6 \\
     \hline
     \end{tabular}
    \label{tab:5}
\end{table}

\begin{table}[]
    \centering
    \caption{Intelligent Iteration distribution : Identifying tampered/bad hardware (Iteration/run = 5 ; $_-$t denotes tampered results)}
    \begin{tabular}{ccccccc}

     \hline
      & & & Approximation ratio (r) \\ 
     \hline
    $t=0.1$ &  $HW$ &  Run 1 & .61 \\
           &          & Run 2 & .62\\
           & $HW_-t$  & Run 1 & .56 \\
           &          & Run 2 & .52 \\
     \hline
     
    $t=0.2$ &  $HW$ &  Run 1 & .62 \\
           &          & Run 2 & .62\\
           & $HW_-t$  & Run 1 & .53 \\
           &          & Run 2 & .5 \\
     \hline
    $t=0.3$ &  $HW$ &  Run 1 & .63 \\
           &          & Run 2 & .62 \\
           & $HW_-t$  & Run 1 &  .51 \\
           &          & Run 2 & .50 \\
     \hline
    $t=0.4$ &  $HW$ &  Run 1 & .62 \\
           &          & Run 2 & .61 \\
           & $HW_-t$  & Run 1 & .49 \\
           &          & Run 2 & .5 \\   
     \hline
    
    $t=0.5$ &  $HW$ & Run 1 & .62 \\
           &          & Run 2 & .66 \\
           & $HW_-t$  & Run 1 & .48 \\
           &          & Run 2 & .49 \\
     \hline
     \end{tabular}
    \label{tab:6}
\end{table}

\subsubsection{Pure quantum workloads}

The Fig. \ref{10} depicts the performance improvement from the 50-50 split countermeasure against adversarial tampering across multiple benchmarks. With t=0.5, we simulate one tamper-free (Fake$_-$montreal) and one tampered hardware (Fake$_-$montreal$_-$ tampered). The user is oblivious of tamper-free hardware. Fig. \ref{6} showed that t=0.5 is so detrimental that the user no longer samples the correct output across all benchmarks. However, a 50-50 split shows a significant improvement ($\approx$ 300$\%$ PM increase on average) for all simulated benchmarks compared to running all shots on the tempered hardware.

Fig. \ref{11} shows the improvement in PM and TVD with \emph{t}, validating the proposed defense for a single benchmark (toffoli$_-$n3) under the same assumption of one tampered and one tamper-free hardware. The proposed 50-50 split results in a significant TVD reduction ($\approx$ 55$\%$ on average) and improvement in PM margin ($\approx$ $125\%$ on average and a maximum of $\approx$ 400$\%$ for t=0.5). The Fig. \ref{12} depicts the improvements in PM and TVD as a function of the percentage of shots run on tamper-free hardware. The benchmark toffoli$_-$n3 is run for 10,000 shots on Fake$_-$montreal (tamper-free hardware) and Fake$_-$montreal$_-$tampered (tampered with t =0.5). We note a $\approx$ 60$\%$ improvement in PM and a 50$\%$ reduction in TVD with the 50-50 split (than when all 10,000 shots are allocated to the tampered hardware). However, as the number of shots assigned to tampered-free hardware increases, so does the PM, with a corresponding drop in TVD. For the 90-10 split, we see a massive improvement in PM of approximately 1900$\%$ and a significant TVD reduction 90$\%$ compared to the tampered hardware. 

A sample simulation of how a user can determine the tamper-free hardware and allocate the majority of the shots to that preferred hardware is shown in the Table \ref{tab:3}. We run two 50-shot runs for two benchmarks, toffoli$_-$n3 (3-qubit benchmark) and adder$_-$n10 (10-qubit benchmark), on two different hardware HW (Fake$_-$montreal) and HW$_-$t(Fake$_-$montreal$_-$tampered). The simulations account for the extent of tampering experienced by HW$_-$t (by varying t from 0.1 to 0.5). We compare the PM, TVD, frequent output, repeatability, and confidence factor (probability) across the two runs for each hardware. 
HW outperforms HW$_-$t for both distinct benchmarks in every way.
Adversarial tampering (even minor tampering e.g., $t=0.1$) along with the temporal variations in quantum hardware leads to the HW$_-$t converging to different outputs for the two different runs for benchmark adder$_-$n10. In contrast, when other factors such as low TVD variation across two runs are considered as well, along with the tamper-free HW producing the same output for both runs, makes it more reliable. As a result, the user can choose to run the remaining shots in HW only. Hence the user can get a performance boost comparable to the 90 to 10 split (as much as 1900$\%$ and  $90\%$ in two chosen
performance metrics) Fig. \ref{12}.


\subsubsection{Hybrid quantum classical workload}

The proposed 50-50 split is also applicable to hybrid-classical algorithms like QAOA. Fig. \ref{18} compares how the objective and the AR converges over 50 iterations using the 50-50 split. Table \ref{tab:5} shows the improvement in AR with t, assuming one tampered (HW$_-$t) and one tamper-free of hardware (HW). We run 25 iterations (50 shots/iteration) on HW (fake$_-$ montreal) out of a total of 50 iterations, extract the parameters $(\gamma, \beta)$ after 25 iterations, and use them as a starting point for parameter optimization in HW$_-$t for another 25 iterations. We observe AR improvement for various levels of tampering. For t=0.5, we report the maximum improvement in AR (15$\%$ on average) across various graph sizes.

Table \ref{tab:6} shows a sample simulation of how a user can determine the tamper-free hardware and allocate the majority of iterations for a hybrid algorithm such as, QAOA to that preferred hardware. We execute two 5-iteration (50 shot/iteration) runs on two different hardware HW (Fake$_-$montreal) and HW$_-$t (Fake$_-$montreal$_-$tampered) for a 4 node graph. The simulations account for the degree of tampering experienced by HW$_-$t (by varying t from 0.1 to 0.5). We compare the approximation ratio between the two runs for each hardware. Hardware with a higher AR is better and more reliable. The user can choose to run the remaining iterations in HW only. As a result, the user will benefit from better performance (upto 15$\%$).

\subsection{Validation of Defense on Real Hardware}

We run a sample experiment on real hardware to validate the effectiveness of the proposed run-adaptive shot splitting heuristic against adversarial tampering. We extend the results of our tampering model from the fake backend simulations for the benchmark toffoli$_-$n3 to mimic tampering in the real IBM device ibmq$_-$manila. Table IV summarizes the experiment's findings. The real ibmq$_-$manila device is represented as manila, and the tampered hardware is manila$_-$t (for which we run our benchmark on actual hardware and tamper by modeling the tampering results obtained from the fake manila that we created). We sweep \emph{t} to mimic adversarial tampering. The user performs two initial runs of 50 shots on each hardware, then compares the PM, TVD, frequent output, repeatability, and confidence factor (probability) across the two runs to determine the superior hardware. Adversarial tampering (even minor tampering with t=0.2) combined with temporal variations in the real quantum hardware causes the tampered hardware to diverge to different outputs for the two different runs. For t=0.2 and above, the tampered hardware manila$_-$t begins to diverge to different correct outputs across runs. The user can now allocate the rest of the shots intelligently on seemingly more reliable hardware manila.

\begin{figure}
    \centering
    \includegraphics[width= 3.25in]{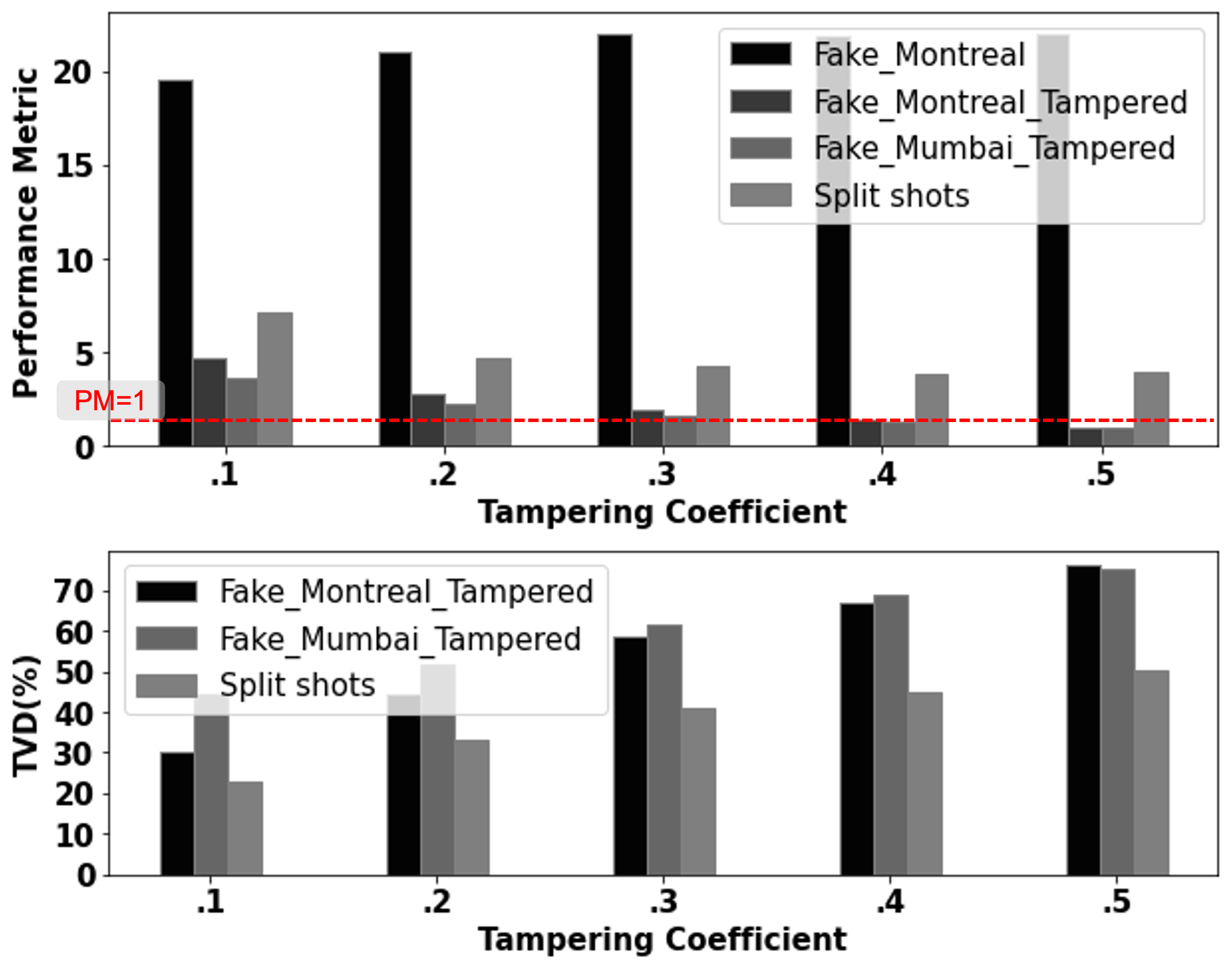}
    \caption{PM improvement and TVD reduction with the 50-50 split heuristic, when user has to choose from 3 different hardware, 2 out of which are tampered (Fake$_-$montreal$_-$tampered, Fake$_-$mumbai$_-$tampered). Benchmark used:  toffoli$_-$n3, number of shots: 10,000.
    }
    \label{14}
\vspace{-4mm}
\end{figure}

\subsection{Generalizing the Proposed Defense}

The proposed heuristics (50-50 split and adaptive-run shots split) provide scalable improvement for a general case where the user must choose among n tampered and one tamper-free hardware (without knowing the idenity of tamper-free hardware). In Fig. \ref{14} we consider 2 tampered (Fake$_-$montreal$_-$tampered, Fake$_-$mumbai$_-$tampered) and 1 tamper-free hardware, with 10,000 shots for the program toffoli$_-$n3. We see PM improvement and TVD reduction when using the 50-50 split heuristic. Similarly, the user can perform two runs with 50 initial shots to determine the reliable hardware and divide the remaining shots accordingly to counter any adversarial tampering.

\subsection{Summary of Defense Analysis}
(a) The proposed 50-50 split effectively mitigates the worst-case tampering scenario where the user originally samples incorrect output. (b) The proposed intelligent run adaptive split enables the user to identify tamper-free hardware. (c) For purely quantum and hybrid workloads, the adaptive split heuristic almost entirely mitigates the proposed adversarial threat. (d) The proposed defense heuristics are applicable to real quantum hardware. (e) The proposed heuristics provide scalable improvement for a general case of n tampered and one tamper-free hardware.

\begin{table}[]
    \centering
    \caption{Validation on IBM hardware (shots/run=50)}
    \begin{tabular}{ccccccc}
     \hline
      & & & PM & TVD(\%) & O/P & Probability  \\ 
     \hline
    t=.1 &  manila & Run 1 & 8.2 &  & 111 & 41/50 \\
           &          & Run 2 & 8 & 10 & 111 & 38/50 \\
           & manila$_-t$  & Run 1 & 2.8 &  & 111 & 21/50 \\
           &          & Run 2 & 2.4 & 24 & 111 & 24/50 \\
     \hline
     
    t=.2 &  manila & Run 1 & 4.4 &  & 111 & 26/50 \\
           &          & Run 2 & 4.8 & 16 & 111 & 29/50 \\
           & manila$_-t$  & Run 1 & .69 &  & 011 & 13/50 \\
           &          & Run 2 & .54 & 28 & 101 & 13/50 \\
     \hline
    t=.3 &  manila & Run 1 & 4.6 & & 111 & 37/50\\
           &          & Run 2 & 5.14 & 8 & 111 & 36/50 \\
           & manila$_-t$  & Run 1 & .53 &  & 011 & 13/50 \\
           &          & Run 2 & 1.1 & 34 & 111 & 12/50 \\
     \hline
    t=.4 &  manila & Run 1 & 8.4 &  & 111 & 42/50\\
           &          & Run 2 & 7.4 & 10 & 111 & 37/50\\
           & manila$_-t$  & Run 1 & 1.2 &  & 111 & 12/50\\
           &          & Run 2 & .41 & 26 & 110 & 12/50\\   
     \hline
    
    t=.5 &  manila & Run 1 & 5.1 &  & 111 & 31/50\\
           &          & Run 2 & 7.2 & 9 & 111 & 36/50 \\
           & manila$_-t$  & Run 1 & .35 &  & 011 & 14/50 \\
           &          & Run 2 & .13 & 31 & 010 & 15/50 \\
     \hline
     \end{tabular}
    \label{tab:4}
\end{table}

\section{Conclusion}

In this paper, we propose an adversarial attack by a less reliable third-party provider. We simulate the impact of adversarial tampering on a variety of purely quantum and hybrid quantum-classical benchmarks. We report an average reduction of 0.12X in the PM and an increase in TVD of 7X across purely quantum workloads for minimally tampered hardware (t=0.1) and an average reduction in AR of 0.8X (t=0.1) and 0.25X (t=.5) for quantum classical workload. We propose distributing the total number of shots available to the user among various hardware options to ensure trustworthy computing using a mix of trusted and untrusted hardware. On average, we note a 30X improvement in PM, a 0.25X reduction in TVD for pure quantum workloads and AR improvement upto 1.5X. Our proposed heuristics (50-50 shot distribution and run-adaptive shot distribution) mitigate the adversary's tampering (random/targeted) efforts, improving the quantum program's resilience.

\section{Acknowledgment}

We thank Dr Rasit Onur Topaloglu (IBM Corporation) for useful discussions. The work is supported in parts by NSF (CNS-1814710, DGE-1821766, CNS-2129675, CCF-2210963, DGE-2113839, ITE-2040667), gifts from Intel, seed grants from Penn State ICDS and Huck Institute of the Life Sciences.

\end{document}